\documentclass[11pt,aps,prd,showpacs,showkeys,superscriptaddress,preprintnumbers,amsmath,amssymb,nofootinbib]{revtex4-2}

\pdfoutput=1 

\usepackage{graphicx}
\usepackage{dcolumn}
\usepackage{bm}
\usepackage{color}
\usepackage{epsfig}
\usepackage{amssymb}
\usepackage{amsmath}
\usepackage{amsfonts}
\usepackage{float}
\usepackage{natbib}
\usepackage{amsmath,amssymb,enumitem}
\usepackage{braket}
\usepackage{fontenc}[T1]
\usepackage[utf8]{inputenc}
\usepackage{multirow}
\usepackage{placeins}
\usepackage{slashed}
\usepackage{braket}
\usepackage{xspace}
 \usepackage{float}
\usepackage[usenames, dvipsnames]{xcolor}
\usepackage{url}
\usepackage[linktocpage=true]{hyperref}
\hypersetup{
   colorlinks   = true,
    linkcolor= blue,
    citecolor= blue,
    urlcolor= blue}

\newcommand{\be}{\begin{equation}}
\newcommand{\ee}{\end{equation}}
\makeatletter

\def\ak{\@ifstar\@@ak\@ak}
\newcommand{\@ak}[1]{\textcolor{ForestGreen}{[\textbf{AK:} #1]}}
\newcommand{\@@ak}[1]{\textcolor{ForestGreen}{#1}}
\makeatother

\begin{document}

\title{\boldmath Magnetic Moments of Hidden-Bottom Pentaquark States}

\author{Halil Mutuk}
\email[Electronic address:~]{hmutuk@omu.edu.tr}
\affiliation{Department of Physics, Faculty of Sciences, Ondokuz Mayis University, 55200 Samsun, Türkiye}
%


\begin{abstract}
\vspace{1cm}
We study systematically magnetic moments of hiddden-bottom pentaquark states with quantum numbers $J^P=\frac{1}{2}^{\pm}$, $J^P=\frac{3}{2}^{\pm}$, and $J^P=\frac{5}{2}^{\pm}$ with molecular, diquark-diquark-antiquark, and diquark-triquark models. The numerical results show that magnetic moments are different within the same model according to same quantum numbers and spin-orbit couplings. The results are also different  when different models are taken into account with the same angular momentum. The magnetic moments encode valuable information about inner structures. We believe that our results may be helpful for experimental studies.

\end{abstract}



\maketitle

\section{Introduction}
\label{sec:intro}
According to quark model, conventional mesons are quark-antiquark states and baryons are three-quark states. Since its introduction in the 60s, this model explained successfully the spectrum of mesons and baryons. This paradigm has changed with the beginning of the millennium. The discoveries of the multiquark states  have opened a new era in the particle physics. These unconventional objects are called exotic hadrons and do not fit the quark model picture due to the some extra ordinary features. The quantum chromodynamics (QCD) which is the theory of strong interactions, allows those unconventional hadrons such as multiquark states $(qq\bar q \bar q$, $qqqq\bar q$, $qqq\bar q \bar q \bar q$, $qqqqqq, \cdots)$, glueballs $(gg, ggg, \cdots)$, and quark-gluon hybrids $(q\bar q g)$. 

The first exotic state $\chi_{c1}(3872)$ which is a four-quark state is observed in 2003 by Belle Collaboration \cite{Belle:2003nnu}. Since then, many exotic hadrons (mostly four-quark states) have been observed in the various experiments. Among these exotic hadrons, pentaquark family has a few members. In 2015, LHCb Collaboration reported two hidden-charm pentaquark states $P_c(4380)$ and $P_c(4450)$ after investigating $J/\psi p$ invariant mass spectrum of $\Lambda_b^0 \to J/\psi p K^-$ process with masses and widths \cite{LHCb:2015yax}:
\begin{eqnarray}
m_{P_c(4380)^+}&=&4380 \pm8 \pm 29~\mathrm{MeV}, \quad \Gamma_{P_c(4380)^+}=205 \pm 18 \pm 86~\mathrm{MeV}, \\
m_{P_c(4450)^+}&=&4449.8 \pm 1.7 \pm 2.5~\mathrm{MeV}, \quad \Gamma_{P_c(4450)^+}= 39 \pm 5 \pm 19~\mathrm{MeV}.
\end{eqnarray}
Four years later after this observation,  LHCb Collaboration reported three new pentaquark states namely  $P_c(4312)$, $P_c(4440)$ and $P_c(4457)$ using updated data \cite{LHCb:2019kea}. The previously reported $P_c(4450)$ was resolved into two narrow states as $P_c(4440)$ and $P_c(4457)$. The masses and widhts of these new pentaquark states are  as follows
\begin{eqnarray}
m_{P_c(4312)^+}&=&4311.9 \pm 0.7^{ +6.8}_{-0.6}~\mathrm{MeV}, \quad \Gamma_{P_c(4312)^+}=9.8 \pm 2.7 ^{ +3.7}_{-4.5}~\mathrm{MeV},\\
m_{P_c(4440)^+}&=&4440.3 \pm 1.3 ^{+ 4.1}_{-4.7}~\mathrm{MeV}, \quad \Gamma_{P_c(4440)^+}= 20.6 \pm 4.9^{+8.7}_{-10.1}~\mathrm{MeV},\\
m_{P_c(4457)^+}&=&4457.3 \pm 0.6 ^{+ 4.1}_{-1.7}~\mathrm{MeV}, \quad \Gamma_{P_c(4457)^+}= 6.4 \pm 2.0^{+5.7}_{-1.9}~\mathrm{MeV}.
\end{eqnarray}
The continuous search on pentaquark states resulted a further discovery. A new pentaquark state with strange quark content namely $P_{cs}(4459)^0$ is reported by the LHCb collaboration via analyzing of $J\psi \Lambda$ invariant mass distribution in $\Xi_b^-\rightarrow J/\psi K^-\Lambda$ decays~\cite{LHCb:2020jpq}. The mass and width for this state is
\begin{equation}
m_{P_{cs}(4459)^0}=4458.8 \pm 2.9^{+4.7}_{-1.1}~\mathrm{MeV}, \quad \Gamma_{P_{cs}(4459)^0} = 17.3 \pm 6.5^{+8.0}_{-5.7}~\mathrm{MeV},
\end{equation}
respectively. Very recently, $P_{cs}(4338)$ state is observed from the amplitude analyses of $B^{-}\rightarrow J/\psi \Lambda \bar{p}$ with mass and width
\begin{equation}
m_{P_{cs}(4338)}=4338.2 \pm 0.7\pm 0.4~\mathrm{MeV}, \quad \Gamma_{P_{cs}(4338)} = 7.0\pm 1.2 \pm 1.3~\mathrm{MeV},
\end{equation}
respectively, \cite{LHCb:2022ogu}.

The observation of the above pentaquark states triggered theoretical and phenomenological studies in order to identify their various properties. Although there are numerous studies regarding spectroscopic and decay parameters of these states, the inner structures and quantum numbers are still unclear. Consecutive studies on these properties provide valuable information about understanding the nonperturbative nature of QCD. In addition to this, such theoretical and phenomenological studies may help experimental studies. 

The aforementioned pentaquark states are investigated by using different models which have different substructures such as diquark-diquark-antiquark states \cite{Lebed:2015tna,Li:2015gta,Maiani:2015vwa,Anisovich:2015cia,Wang:2015ava,Wang:2015epa,Wang:2015ixb,Ghosh:2015ksa,Wang:2015wsa,Zhang:2017mmw,Wang:2019got,Wang:2020rdh,Ali:2020vee,Wang:2016dzu,Wang:2020eep}, diquark-triquark states \cite{Zhu:2015bba,Lebed:2015tna,Wang:2016dzu}, meson-baryon molecular states \cite{Chen:2015loa,Chen:2015moa,He:2015cea,Meissner:2015mza,Roca:2015dva,Wang:2018waa,Wang:2019hyc,Chen:2020uif,Peng:2020hql,Chen:2020kco,Chen:2020opr,Wang:2021itn,Wang:2023eng,Li:2024wxr,Li:2024jlq} and topological soliton model \cite{Scoccola:2015nia}.

Doubly-heavy tetraquarks $(QQ\bar{q}\bar{q})$ should be stable when the ratio $m_{QQ}/m_{\bar{q}\bar{q}}$ is sufficiently large \cite{Carlson:1987hh,Manohar:1992nd}. Before and after the observation of $T_{cc}^+$  \cite{LHCb:2021vvq,LHCb:2021auc} which is a tetraquark state of $cc\bar{u}\bar{d}$ quark content, lattice QCD studies found some evidences about doubly-heavy tetraquarks with $b$ quark content \cite{Francis:2016hui,Junnarkar:2018twb,Leskovec:2019ioa,Mohanta:2020eed,Alexandrou:2023cqg}. Same logic can be applied to hidden-charm pentaquark states by replacing $c$ quarks with $b$ quarks. Hidden-charm pentaquark states were studied in the literature with various methods. Here we collect some studies that are devoted to hidden-bottom pentaquarks. In addition to $P_c(4380)^+$  state, Ref. \cite{Shimizu:2016rrd} contains mass predictions of hidden-bottom pentaquark states with a mass around $11.08-11.11$ GeV and $J^P=3/2^-$ quantum number is favored. Furthermore, in the same work it was mentioned that there may exist some loosely-bound molecular pentaquark states with heavy quark contents of $c\bar{b}$, $b\bar{c}$ and $b\bar{b}$. Such pentaquark masses were calculated as $M_{\bar{c}c}=4678 ~\text{MeV}$, $M_{\bar{b}c}=7789 ~\text{MeV}$ and $M_{\bar{b}b}=11198 ~\text{MeV}$ in Ref. \cite{Liu:2017xzo} by using a variant of D4-D8 brane model \cite{Sakai:2004cn}. Masses and residues of spin 3/2 and 5/2 hidden-bottom pentaquark states with both negative and positive parities are calculated via QCD sum rule formalism \cite{Azizi:2017bgs}. The obtained mass values are $M=10.93^{+0.82}_{-0.85}~\text{GeV}$ for $J^P=\frac{3}{2}^+$, $M=10.96^{+0.84}_{-0.88}~\text{GeV}$ for $J^P=\frac{3}{2}^-$, $M=11.94^{+0.84}_{-0.82}~\text{GeV}$ for $J^P=\frac{5}{2}^+$, and $M=10.98^{+0.82}_{-0.82}~\text{GeV}$ for $J^P=\frac{5}{2}^-$. Ref. \cite{Yang:2018oqd} prompted chiral quark model to the hidden-bottom pentaquark states with spin-parity quantum numbers  $J^P=\frac{1}{2}^{\pm}$, $J^P=\frac{3}{2}^{\pm}$ and $J^P=\frac{5}{2}^{\pm}$ and in the $\frac{1}{2}$ and $\frac{3}{2}$ isospin sectors. They found no bound state candidates for positive parity hidden-bottom pentaquark states. In Ref. \cite{Zhu:2020vto}, possible hidden-bottom pentaquark states from coupled-channel $\Sigma^{(*)}_b B^{(*)}-\Lambda_b B^{(*)}$ interaction are investigated by using quasipotential Bethe-Salpeter equation approach. They found seven molecular states can be produced from the interactions. Five-quark systems composed of $qqs\bar{Q}Q$ configuration with ($q = u$ or $d$, $Q=b$ or $c$) are studied in the chiral quark model \cite{Zhang:2020cdi}. They observed that color-octet structure yielded more bounding energy than color-singlet structure. In Ref. \cite{Paryev:2023uhl}, production of non-strange hidden-bottom pentaquark states near threshold $\Upsilon(1S)$ meson photoproduction is studied. Hidden-heavy pentaquark and $P_{cs}$ states are studied in the framework of MIT bag model \cite{Zhang:2023teh}. Apart from $P_{cs}$ states, hadron properties and decay channels of $P_{bss}$ and $P_{bsss}$ are obtained. Masses, magnetic moments and partial widths of hidden-bottom pentaquarks are studied by using extension of the Gursey-Radicati mass formula,  effective mass  and screened charge schemes \cite{Sharma:2024ern}. 

In the light of experimental and theoretical developments, it is natural to expect bottom analogues of the observed hidden-charm penta quark states. Indeed, historically the observation of hadrons with $c$ quarks is followed by similar structures with $b$ quarks content. In this path, LHCb Collaboration sparked a light for pentaquark states containing a single $b$ quark which decay weakly via the $b\to c\overline{c}s$ transition in the $J/\psi K^+\pi^- p$, $J/\psi K^- \pi^- p$, $J/\psi K^- \pi^+ p$,  $J/\psi \phi p$ final states \cite{LHCb:2017wmj}. They reported about similar explorations for hidden-bottom pentaquark states should be expected in the near future. 

The literature review reflects that the majority of studies has concentrated on spectroscopic (such as mass) and decay (such as pole residue) parameters of hidden-bottom pentaquark states. Such properties may be insufficient to elucidate the inner structures of these states. Therefore further studies that cover electromagnetic properties, radiative and weak decays are required. Unveiling the nature of the unconventional hadronic states is a major concern of high energy physics. Physical quantities associated with the electromagnetic properties are valuable parameters. In this respect, magnetic moments may provide valuable insights about the internal structure of both conventional and unconventional hadrons. They present information about charge distributions and shape of the hadrons. Magnetic moment is also an excellent probe for quark-gluon dynamics inside the hadron since it is the leading-order response of a bound state to a weak external magnetic field \cite{Ozdem:2023rkx}. 

In this present work, we systematically study magnetic moments of hidden-bottom pentaquarks using diquark-diquark-antiquark scheme, diquark-tiruqark scheme and meson-baryon molecule scheme. In Section \ref{sec:models}, we present the models that are used in this study. Section \ref{sec:formalism} is related to wave functions of different models. Sections \ref{sec:molecular}, \ref{sec:dda}, and \ref{sec:dda} are devoted to  magnetic moment results of molecular, diquark-diquark-antiquark, and diquark-triquark respectively. Last two sections are reserved for discussion and conclusion.

\section{Models}
\label{sec:models}
In this section, we categorize three different models that are used to study magnetic moments of hidden-bottom pentaquark states.
Pentaquark states can be decomposed into two or three clusters in quark level and color configurations. Furthermore, these clusters should combine in a way to become color neutral. 

The color interaction plays important role when determining the configurations of clusters and consequently pentaquark states. In quark potential model, color factor enters through the Coulombic part. The color factor $f$ in the quark-antiquark color interaction is
\begin{equation}
\kappa=-\frac{1}{4}\sum_{i<j,a=1}^8 \lambda_i^a \lambda_j^a,
\end{equation}
where $\lambda_a^i$ is the color generator of the ith quark. Then the corresponding Coulombic potential is
\begin{equation}
V_{q\bar{q}}(r)\simeq   \kappa \frac{\alpha_s}{r}.
\end{equation}
The quark-quark color factor is given by
\begin{equation}
\tilde{\kappa}=\frac{1}{4}\sum_{i<j,a=1}^8 \lambda_i^a \lambda_j^a,
\end{equation}
and the Coulombic potential is
\begin{equation}
V_{qq}(r)\simeq  \tilde{\kappa} \frac{\alpha_s}{r}.
\end{equation}

The SU(3) color symmetry of QCD implies that, when we combine a quark and an antiquark in the fundamental color representation, we obtain $\vert q \bar{q} \rangle:  3  \otimes  \bar{3}= 1 \oplus 8 $. This representation gives the color factor for the color singlet as $\kappa=-4/3$ of the quark-antiquark system. When we combine two quarks in the fundamental color representation, it reduces to $\vert q q \rangle:  3  \otimes  3= \bar{3} \oplus 6 $, a color antitriplet  $\bar{3}$ and a color sextet $6$. The antitriplet state has a color factor $\kappa=-2/3$ which is attractive whereas the sextet state  has a color factor $\kappa=+1/3$ which is repulsive. Since an attractive potential is responsible for binding quarks together, color factors are important to determine the configurations for hadronic states. Table \ref{tab:colorfactors} presents the color factor values.

\begin{table}[h]
    \renewcommand{\arraystretch}{1.25}
    \centering
\begin{tabular}{|c|c|c|c|}
    \hline
      State  & Quark configuration & Color configuration & Color factor \\ \hline\hline
 \multirow{2 }{*} {Meson}     &\multirow{2}{*} {$\vert q \bar{q} \rangle$}   & \multirow{2}{*} {$3_{c} \otimes \bar{3}_{c}=1_{c} \oplus 8_{c}$}    & $1_{c} \oplus 8_{c}$   \\
&&       & $-\frac{4}{3} \quad \frac{1}{6}$   \\ \hline

 \multirow{2 }{*} {Diquark}  &\multirow{2}{*} {$\vert q q \rangle$}   & \multirow{2}{*} {$3_{c} \otimes 3_{c}=6_{c} \oplus \bar{3}_{c}$}    & $6_{c} \oplus \bar{3}_{c}$    \\
&&       & $ \frac{1}{3} \quad -\frac{2}{3}$   \\ \hline

 \multirow{2 }{*} {Baryon}     &\multirow{2}{*} {$\vert q q  q\rangle$}   & \multirow{2}{*} {$3_{c} \otimes 3_{c} \otimes 3_{c}=1_{c} \oplus 8_{1c} \oplus 8_{2c} \oplus 10_{c} $}    & $1_{c} \oplus 8_{1c} \oplus 8_{2c} \oplus 10_{c} $    \\
&&       & $ -2$ \quad $-\frac{1}{2}$ \quad $-\frac{1}{2}$ \quad 1   \\ \hline

\multirow{2 }{*} {Triquark}     &\multirow{2}{*} {$\vert q q  \bar{q} \rangle$}   & \multirow{2}{*} {$3_{c} \otimes 3_{c} \otimes \bar{3}_{c} =3_{1c} \oplus 3_{2c} \oplus \bar{6}_{c} \oplus 15_{c} $}    & $3_{1c} \oplus 3_{2c} \oplus \bar{6}_{c} \oplus 15_{c} $    \\
&&       &  $-\frac{4}{3}$ \ \  $- \frac{4}{3}$ \ \ $- \frac{1}{3}$ \ \ $2$   \\
 \hline
 \hline
              
\end{tabular}
      \caption{
Color factors of hadronic configurations. Color factors are given beneath the corresponding color structures.}
\label{tab:colorfactors}
\end{table}

Since color confinement restricts hadrons to be color singlets, one can classify pentaquarks into three categories:

\begin{enumerate}
\item Molecular model

Molecular states of pentaquarks are assumed to be composed of mesons and baryons. Each cluster in the molecular model tends to be color singlets. Looking color factors of mesons, it can be seen that color factor of $1_c$ is negative and while it is positive for $8_c$. In other words $\kappa_{1_c} < \kappa_{8_c}$ so that a singlet state is more likely to form than octet state. Looking color factors of baryons, color factors of  $(1_c, 8_{1c}, 8_{2c})$ color representations are negative whilst $10_c$ color representation has positive color factor. As mentioned before, negative values of color factor are attractive and positive values are repulsive. The color factors are ordered as $\kappa_{1_c} < \kappa_{8_{1c}} = \kappa_{8_{1c}} < \kappa_{10_c}$. This implies that it is easier to form a singlet state. As a result, there are two configurations for molecular model: $(b\bar{b}) (q_1 q_2 q_3)$ and $(\bar{b} q_1) (b q_2 q_3)$, where $q$ denotes light quarks, $u,d,s$. 

\item Diquark-diquark-antiquark model

The diquark cluster has color factor of $\kappa=-\frac{2}{3}$ for $\bar{3}_c$ color representation and $\kappa=\frac{1}{3}$ for $6_c$ color representation. The diquark prefers to form $\bar{3}_c$, therefore each diquark cluster recasts as $ \bar{3}_c \otimes \bar{3}_c $ to form $3_c$ color representation. The antiquark has also $\bar{3}_c$ to form color neutral state. This consideration brings us to have $\bar{3}_c ({\text{Diquark}}) \otimes \bar{3}_c (\text{Diquark}) \otimes  \bar{3}_c (\text{Antiquark})$ configuration. Hence, the pentaquark configuration $(b q_1) (q_2 q_3) (\bar{b})$ is the nontrivial one according to diquark-diquark-antiquark model. 

\item Diquark-triquark model

As in the diquark-diquark-antiquark model, diquark prefers $\bar{3}_c$ color representation. Triquark has two quarks and one antiquark which distinguish it from a conventional baryon. Looking color factors of triquark, it can be seen that $\kappa_{3_{1_c}} = \kappa_{3_{2c}} < \kappa_{\bar{6}_c}<\kappa_{\bar{15}_c} $. It is more likely to bind in $3_c$ color representation. As a result we have $3_c ({\text{Triquark})} \otimes \bar{3}_c ({\text{Diquark})} $ to recast in a color neutral state. Thus, the pentaquark configurations related to diquark-triquark model are $(bq_1) (\bar{b}q_2q_3)$ and $(\bar{b}b q_1) (q_2 q_3) $.

\end{enumerate}

Some pentaquark configurations may be omitted due to some kinematic and dynamical reasons. The clusters $(b \bar{b})$ and $(\bar{b}q_1)$ tend to the color singlet states. $(b \bar{b})$ is the botomonium and $\bar{b}q_1$ is a heavy-light meson. There is no strong attraction between a bottomonium and a light baryon $(q_1 q_2 q_3)$. As a result, it is hard to form a loosely bound molecular state composed of $(b \bar{b})$ and $(q_1 q_2 q_3)$. Furthermore, as pointed out in Ref. \cite{LHCb:2019kea}, the separation of charm $c$ and anti-charm $\bar{c}$ into distinct confinement volumes provides a natural suppression mechanism for the pentaquark widths. Since the separation of bottom $b$ and anti-bottom $\bar{b}$ in the botomonium is less then the charmonium, such suppresion may occur in the hidden-bottom pentaquark states. Therefore, we do not consider $(b \bar{b})(q_1 q_2 q_3)$ configuration in molecular model and $(\bar{b}b q_1) (q_2 q_3)$ configuration in diquark-triquark model.

The aforementioned models presented good description for the experimentally observed hidden-charm pentaquark states in terms of mass and decay properties. However, mass and decay width alone cannot distinguish inner structures of these pentaquark states. Magnetic moment is an important dynamical observable for a hadron which reflects indications about charge distributions, shape and magnetization. Although hidden-bottom pentaquark states are not experimentally observed, any study regarding magnetic moment may present underlying reasons for distinguishing models and identify inner structure.

\section{Wave Function of Hidden-Bottom Pentaquark States}
\label{sec:formalism}
 There are four degrees of freedom in the quark level: color, spin, flavor and space. Accordingly one can write the wavefunction of a hadronic state as
\begin{equation}
\Psi_{\text{Wavefunction}}=\psi_{\text{flavor}} \psi_{\text{spin}} \psi_{\text{color}} \psi_{\text{space}}.
\end{equation}
The overall wave function should be antisymmetric due to the Fermi statistics. We construct hidden-bottom pentaquark states in the $SU(3)_f$ frame. The $q_2 q_3$ forms $\bar{3}_f$ and $6_f$ flavor representation with the total spin $S=0$ and $S=1$, respectively. If $q_2 q_3$ forms in $6_f$, then $q_1$ combines with $q_2 q_3$ to form $6_f \otimes 3_f =10_f \oplus 8_{1f}$. If $q_2 q_3$ forms in $\bar{3}_f$, then $q_1$ combines with $q_2 q_3$ to form $\bar{3}_f \otimes 3_f =8_{2f} \oplus 1_f$. After inserting $b$ quark and $\bar{b}$ antiquark together with the Clebsch-Gordan coefficients, we obtain the flavor wave function in the molecular model with the $(\bar{b} q_1) (b q_2 q_3)$ configuration. The same method can be applied to diquark-diquark-antiquark $(b q_1) (q_2 q_3) (\bar{b})$ and diquark-triquark $(bq_1) (\bar{b}q_2q_3)$ configurations. We list the wave functions in Table \ref{tab:wavefunction}.
 
\begin{table}[h]
    \centering
    \renewcommand{\arraystretch}{1.25}
    \begin{tabular}{|c|c|c|c|c|}
        \hline
        Model  &  Multiplet  &  $(I, I_3)$  &  Wave function  \\\hline\hline
        Molecular   \,&\,  $10_f$ \ \,&\, $(\frac{3}{2},-\frac{1}{2})$ \,&\,   $\sqrt{\frac{2}{3}}({\bar b}d)(b\{ud\})+\sqrt{\frac{1}{3}}({\bar b}u)(b\{dd\})$  \\   
  \,&\,  $8_{1f}$ \ \,&\, $(\frac{1}{2},-\frac{1}{2})$ \,&\,   $\sqrt{\frac{1}{3}}({\bar b}d)(b\{ud\})-\sqrt{\frac{2}{3}}({\bar b}u)(b\{dd\})$     \\
 \,&\,  $8_{2f}$ \ \,&\, $(\frac{1}{2},-\frac{1}{2})$ \,&\,   $({\bar b}d)(b[ud])$      
\\ \hline \hline

 Diquark-diquark-antiquark\,&\,  $10_f$ \ \,&\, $(\frac{3}{2},-\frac{1}{2})$\,&\,  $\sqrt{\frac{2}{3}}({b}d)\{ud\}{\bar b}+\sqrt{\frac{1}{3}}({b}u)\{dd\}{\bar b}$ \\ 
  \,&\,  $8_{1f}$ \ \,&\, $(\frac{1}{2},-\frac{1}{2})$\,&\,  $\sqrt{\frac{1}{3}}(bd)\{ud\}{\bar b}-\sqrt{\frac{2}{3}}({b}u)\{dd\}{\bar b}$ \\ 
 \,&\,  $8_{2f}$ \ \,&\, $(\frac{1}{2},-\frac{1}{2})$\,&\,  $(bd)[ud]{\bar b}$ \\  \hline \hline

Diquark-triquark\,&\,  $10_f$ \ \,&\, $(\frac{3}{2},-\frac{1}{2})$\,&\, $\sqrt{\frac{2}{3}}({b}d)(\bar b\{ud\})+\sqrt{\frac{1}{3}}({b}u)(\bar b\{dd\})$ \\ 
  \,&\,  $8_{1f}$ \ \,&\, $(\frac{1}{2},-\frac{1}{2})$\,&\,  $\sqrt{\frac{1}{3}}({b}d)(\bar b\{ud\})-\sqrt{\frac{2}{3}}({b}u)(\bar b\{dd\})$ \\ 
 \,&\,  $8_{2f}$ \ \,&\, $(\frac{1}{2},-\frac{1}{2})$\,&\,  $({b}d)(\bar b[ud])$ \\

  \hline \hline
      \end{tabular}
        \caption{The flavor wave functions of the hidden-bottom pentaquark states in different models. Here, $\{q_1q_2\}=\frac{1}{\sqrt 2}(q_1q_2+q_2q_1)$ and $[q_1q_2]=\frac{1}{\sqrt 2}(q_1q_2-q_2q_1)$. $I$ and $I_3$ are the isospin and its third component, respectively.}
          \label{tab:wavefunction}
\end{table}

In this present paper, we obtain magnetic moments of hidden-bottom pentaquark states with strangeness zero. In other words, the pentaquark states of this work do not contain $s$ quark. 

\section{Magnetic Moments of Hidden-Bottom Pentaquark States in Molecular Model}
\label{sec:molecular}
The magnetic moment of the pentaquark state with the molecular configuration $(\bar{b} q_1) (b q_2 q_3)$ can be written as
\begin{equation}
\hat{\mu}  = \ \hat{\mu}_{B}+\hat{\mu}_{B}+\hat{\mu}_{l},
\end{equation}
where the subscripts $B$ and $M$ represent the baryon and meson, respectively, and $l$ is the orbital excitation between the meson and baryon. We assume that the orbital excitation occurs between the meson and baryon and do not consider the orbital excitation in the bound state. The expressions for the magnetic moments of meson, baryon and orbital excitation are given as
\begin{eqnarray}
		\hat{\mu}_{B}  &=& \sum_{i=1}^{3} \mu_{i}g_{i}\hat{S}_{i},\\
		\hat{\mu}_{M}  &=& \sum_{i=1}^{2} \mu_{i}g_{i}\hat{S}_{i},\\
		\hat{\mu}_{l}  = \mu_{l}\hat{l} &=& \frac{M_M\mu_B+M_B\mu_M}{M_M+M_B}\hat{l}.
	\end{eqnarray}
where $g_{i}$ is the Lande factor, $M_M$ and $M_{B}$ are the masses of meson and baryon, respectively. The magnetic moment expression for $(\bar{b} q_1) (b q_2 q_3)$ molecular configuration is
\begin{eqnarray}
		\mu  &=& \langle\ \psi_{Pentaquark}\ |\ \hat{\mu}_B+\hat{\mu}_M+\hat{\mu}_{l}\ |\ \psi_{Pentaquark}\  \rangle\nonumber\\
		&=&
		\sum_{SS_z,ll_z}\ \langle\ SS_z,ll_z|JJ_z\ \rangle^{2}  \left \{ \mu_{l} l_z 	+
		\sum_{S^\prime_B,S^\prime_M}\ \langle\ S_B S^\prime_{B},S_M S^\prime_{M}|SS_z\ \rangle^{2} \Bigg [
		S^\prime_{M}\bigg(\mu_{\bar{b}} + \mu_{q_1}\bigg )\nonumber\right.\\
		&+&\left.
		\sum_{S^\prime_{b}}\ \langle\ S_b S^\prime_{b},S_{D} S^\prime_{B}-S^\prime_{b}|S_{B} S^\prime_{B}\rangle^{2}\bigg(g\mu_{b}S^\prime_{b}+(S^\prime_{B}-S^\prime_{b})(\mu_{q_{2}}+\mu_{q_{3}})\bigg )
		\Bigg ]\right \}.           
\end{eqnarray}
Here  $\psi$ represents the wave function listed in Table  \ref{tab:wavefunction}. $S_M$, $S_B$, $S_D$ are the meson, baryon, and the diquark spin inside the baryon, respectively, and $S^\prime$ is the third spin component. 

We use the following constituent quark masses \cite{Lichtenberg:1976fi}:
\begin{eqnarray}
m_u \ =\ m_d \ =\  336\ \mbox{MeV}, \ 	m_s \ =\ 540\ \mbox{MeV},\  m_b \ =\ 4700\ \mbox{MeV}, \nonumber
\end{eqnarray}
which lead to the magnetic moments of the  octet baryons in rough agreement with the experimental values listed in Table \ref{tab:magmomexp}.

\begin{table}[h]
    \centering
    \renewcommand{\arraystretch}{1.25}
    \begin{tabular}{|c|c|c|c|c|c|c|}
        \hline
        Baryon &  Magnetic moment  & This work &  Experiment \cite{Workman:2022ynf}   \\\hline\hline
$p$  \,&\,  $\frac{4}{3}\mu_u-\frac{1}{3}\mu_d$ \,&\, $2.792$ \,&\, $2.793$  \\
$n$  \,&\,  $\frac{4}{3}\mu_d-\frac{1}{3}\mu_u$ \,&\, $-1.861$ \,&\, $-1.913$  \\
$\Lambda$  \,&\,  $\mu_s$ \,&\, $-0.579$ \,&\, $-0.613 \pm 0.006$  \\   
$\Sigma^+$  \,&\,   $\frac{4}{3}\mu_u-\frac{1}{3}\mu_s$ \,&\, $2.674$ \,&\, $2.460 \pm 0.006$  \\   
$\Sigma^-$  \,&\,   $\frac{4}{3}\mu_d-\frac{1}{3}\mu_s$ \,&\, $-1.048$ \,&\, $-1.160 \pm 0.025$  \\ 
$\Xi^0$  \,&\,   $\frac{4}{3}\mu_s-\frac{1}{3}\mu_u$ \,&\, $-1.392$ \,&\, $-1.250 \pm 0.014$  \\ 
$\Xi^-$  \,&\,   $\frac{4}{3}\mu_s-\frac{1}{3}\mu_d$ \,&\, $-0.462$ \,&\, $-0.651\pm 0.0025$  \\ 
$\Omega^-$  \,&\,   $3\mu_s$ \,&\, $-1.737$ \,&\, $-2.020\pm 0.05$  \\ 
 \hline \hline
    \end{tabular}
        \caption{Magnetic moments of the octet baryons. The results are listed in unit of nuclear magneton $\mu_N$.}
          \label{tab:magmomexp}
\end{table}
Consider the $8 _{1f} $ flavor representation of $\sqrt{\frac{1}{3}}(b\{ud\})({\bar{b}}d) -\sqrt{\frac{2}{3}}(b\{dd\})({\bar{b}}u)$ as an example. Let's take $J^{P}=\frac{1}{2}^{-}({\frac{1}{2}}^{+}\otimes0^{-}\otimes0^{+})$ where $J_{B}^{P_{B}}\otimes J_{M}^{P_{M}}\otimes J_{l}^{P_{l}}$ correspond to the angular momentum and parity of baryon, meson and orbital, respectively. We obtain the  magnetic moment expression as 
\begin{align}
		\mu&=\left\langle \psi_{Pentaquark} \left |g_{M}\mu_{M}{\vec{S}_{M}}+g_{B}\mu_{B}{\vec{S}}_{B}+{\mu_l}\vec{l}\,\,\right|\psi_{Pentaquark}\right\rangle \nonumber\\
		& =
		\langle 00,\frac{1}{2}\frac{1}{2} |\frac{1}{2}\frac{1}{2}\rangle^{2}\Big [
		\langle \frac{1}{2}\frac{1}{2},0 0 |\frac{1}{2}\frac{1}{2}\rangle^{2}
		\big (\langle \frac{1}{2}\frac{1}{2},1 0 |\frac{1}{2}\frac{1}{2}\rangle^{2}		\mu_{b}
		\nonumber\\ 
		& \ +	\langle \frac{1}{2}-\frac{1}{2},1 1 |\frac{1}{2}\frac{1}{2}\rangle^{2} (-\mu_{b}+\frac{1}{3}\mu_{u}+\frac{5}{3}\mu_{d})\big)	\Big ]
		\nonumber\\ 
		& =
		\frac{1}{9} (2\mu_{u}+10\mu_{d}-3\mu_{b} ).
\end{align}

The magnetic moments of hidden-bottom pentaquark states in molecular model are listed in Table \ref{tab:magmom81f} for $8_{1f}$, Table \ref{tab:magmom82f} for $8_{2f}$, and Table \ref{tab:magmom10f} for $10_{f}$ representations, respectively.


\begin{table}[H]
    \centering
    \renewcommand{\arraystretch}{1.25}
    \begin{tabular}{|c|c|c|c|c|c|c|}
        \hline
        $J^P$ &  $^{2s+1}L_j$  & \quad $J_{B}^{P_{B}} \otimes J_{M}^{P_{M}}\otimes J_{l}^{P_{l}}$ \quad &  \quad Magnetic moment   \quad \\\hline\hline
\quad $\frac{1}{2}^-$ \quad  \,&\,  $\quad ^2S_{\frac{1}{2}}$ \quad \,&\, ${\frac{1}{2}}^{+}\otimes 0^{-} \otimes 0^{+}$ \,&\, $-0.598$  \\
\,&\, \,&\, ${\frac{1}{2}}^{+} \otimes 1^{-} \otimes 0^{+}$ \,&\, $0.864$\\
\,&\, \,&\, ${\frac{3}{2}}^{+} \otimes 1^{-} \otimes 0^{+}$ \,&\, $-0.886$\\ \hline

$\frac{3}{2}^-$  \,&\,  $^4S_{\frac{3}{2}}$ \,&\, ${\frac{1}{2}}^{+}\otimes 1^{+} \otimes 0^{-}$ \,&\, $0.399$  \\
 \,&\, \,&\, ${\frac{3}{2}}^{+} \otimes 0^{-} \otimes 0^{+}$ \,&\, $-0.997$  \\
 \,&\, \,&\, ${\frac{3}{2}}^{+} \otimes 1^{+} \otimes 0^{-}$ \,&\, $0.333$ \\
\hline

$\frac{5}{2}^-$  \,&\,  $^6S_{\frac{5}{2}}$ \,&\, ${\frac{3}{2}}^{+}\otimes 1^{-} \otimes 0^{+}$ \,&\, $0.237$  \\
\hline

$\frac{1}{2}^+$  \,&\,  $^2P_{\frac{1}{2}}$ \,&\, ${\frac{1}{2}}^{+} \otimes 0^{-} \otimes 1^{-}$ \,&\, $-0.756$  \\ 
 \,&\,  \,&\, ${(\frac{1}{2}}^{+} \otimes 1^{-})_{\frac{1}{2}} \otimes 1^{-}$ \,&\, $-1.243$  \\ 
\hline
 \,&\,  $^4P_{\frac{1}{2}}$ \,&\, $({\frac{1}{2}}^{+} \otimes 1^{-})_{\frac{3}{2}} \otimes 1^{-}$ \,&\, $0.699$  \\ 
 \,&\,   \,&\, ${\frac{3}{2}}^{+} \otimes 0^{-} \otimes 1^{-}$ \,&\, $-0.041$  \\ 
\hline
 \,&\,  $^2P_{\frac{1}{2}}$ \,&\, ${(\frac{3}{2}}^{+} \otimes 1^{-})_{\frac{1}{2}} \otimes 1^{-}$ \,&\, $1.152$  \\ \hline
 \,&\,  $^4P_{\frac{1}{2}}$ \,&\, ${(\frac{3}{2}}^{+} \otimes 1^{-})_{\frac{3}{2}} \otimes 1^{-}$ \,&\, $0.293$  \\ \hline

$\frac{3}{2}^+$  \,&\,  $^2P_{\frac{3}{2}}$ \,&\, ${\frac{1}{2}}^{+} \otimes 0^{-} \otimes 1^{-}$ \,&\, $-2.031$  \\ 
 \,&\, \,&\, ${(\frac{1}{2}}^{+} \otimes 1^{-})_{\frac{1}{2}} \otimes 1^{-}$ \,&\, $-0.568$  \\ \hline
 \,&\,  $^4P_{\frac{3}{2}}$ \,&\, ${(\frac{1}{2}}^{+} \otimes 1^{-})_{\frac{3}{2}} \otimes 1^{-}$ \,&\, $-0.280$  \\ 
\,&\, \,&\, ${\frac{3}{2}}^{+} \otimes 0^{-} \otimes 1^{-}$ \,&\, $-1.304$ \\ 
\hline
  \,&\,  $^2P_{\frac{3}{2}}$ \,&\, ${(\frac{3}{2}}^{+} \otimes 1^{-})_{\frac{1}{2}}  \otimes 1^{-}$ \,&\, $0.399$  \\ 
\hline
  \,&\,  $^4P_{\frac{3}{2}}$ \,&\, ${(\frac{3}{2}}^{+} \otimes 1^{-})_{\frac{3}{2}}  \otimes 1^{-}$ \,&\, $0.270$  \\ 
\hline
  \,&\,  $^6P_{\frac{3}{2}}$ \,&\, ${(\frac{3}{2}}^{+} \otimes 1^{-})_{\frac{5}{2}}  \otimes 1^{-}$ \,&\, $0.860$  \\ 
\hline

$\frac{5}{2}^+$  \,&\,  $^4P_{\frac{5}{2}}$ \,&\, ${\frac{1}{2}}^{+} \otimes 1^{-} \otimes 1^{-}$ \,&\, $-1.034$  \\ 
 \,&\,  \,&\, ${\frac{3}{2}}^{+} \otimes 0^{-} \otimes 1^{-}$ \,&\, $-2.430$  \\ 
 \,&\,  \,&\, ${(\frac{3}{2}}^{+} \otimes 1^{-})_{\frac{3}{2}} \otimes 1^{-}$ \,&\, $-1.765$  \\
\hline
 \,&\,  $^6P_{\frac{5}{2}}$ \,&\, ${(\frac{3}{2}}^{+} \otimes 1^{-})_{\frac{5}{2}} \otimes 1^{-}$ \,&\, $-0.409$  \\ \hline \hline
    \end{tabular}
        \caption{Magnetic moments of the $8_{1f}$ representation of the molecular hidden-bottom  pentaquark state.  $J_{B}^{P_{B}}$  corresponds to the angular momentum and parity of baryon,  $J_{M}^{P_{M}}$ is for meson and  $J_{l}^{P_{l}}$ is for orbital excitation between meson and baryon. The results are listed in unit of nuclear magneton $\mu_N$.}
 \label{tab:magmom81f}
\end{table}


\begin{table}[H]
    \centering
    \renewcommand{\arraystretch}{1.25}
    \begin{tabular}{|c|c|c|c|c|c|c|}
        \hline
     \quad   $J^P$ \quad & \quad $^{2s+1}L_j$ \quad  & \quad $J_{B}^{P_{B}} \otimes J_{M}^{P_{M}}\otimes J_{l}^{P_{l}}$ \quad &  \quad Magnetic moment  \quad   \\\hline\hline
$\frac{1}{2}^-$  \,&\,  $^2S_{\frac{1}{2}}$ \,&\, ${\frac{1}{2}}^{+}\otimes 0^{-} \otimes 0^{+}$ \,&\, $-0.066$  \\
\,&\, \,&\, ${\frac{1}{2}}^{+} \otimes 1^{-} \otimes 0^{+}$ \,&\, $-0.532$\\
\hline

$\frac{3}{2}^-$  \,&\,  $^4S_{\frac{3}{2}}$ \,&\, ${\frac{1}{2}}^{+}\otimes 1^{-} \otimes 0^{+}$ \,&\, $-0.930$  \\
\hline

$\frac{1}{2}^+$  \,&\,  $^2P_{\frac{1}{2}}$ \,&\, ${\frac{1}{2}}^{+} \otimes 0^{-} \otimes 1^{-}$ \,&\, $-0.218$  \\ 
 \,&\,  \,&\, ${(\frac{1}{2}}^{+} \otimes 1^{-})_{\frac{1}{2}} \otimes 1^{-}$ \,&\, $-0.055$  \\ 
\hline
 \,&\,  $^4P_{\frac{1}{2}}$ \,&\, $({\frac{1}{2}}^{+} \otimes 1^{-})_{\frac{3}{2}} \otimes 1^{-}$ \,&\, $-0.397$  \\ 
\hline

$\frac{3}{2}^+$  \,&\,  $^2P_{\frac{3}{2}}$ \,&\, ${\frac{1}{2}}^{+} \otimes 0^{-} \otimes 1^{-}$ \,&\, $-0.427$  \\ 
 \,&\, \,&\, ${(\frac{1}{2}}^{+} \otimes 1^{-})_{\frac{1}{2}} \otimes 1^{-}$ \,&\, $-0.914$  \\ \hline
 \,&\,  $^4P_{\frac{3}{2}}$ \,&\, ${(\frac{1}{2}}^{+} \otimes 1^{-})_{\frac{3}{2}} \otimes 1^{-}$ \,&\, $-0.826$  \\ 
\hline

$\frac{5}{2}^+$  \,&\,  $^4P_{\frac{5}{2}}$ \,&\, ${\frac{1}{2}}^{+} \otimes 1^{-} \otimes 1^{-}$ \,&\, $-1.291$  \\ 

 \hline \hline
    \end{tabular}
        \caption{Magnetic moments of the $8_{2f}$ representation of the molecular hidden-bottom  pentaquark state.  $J_{B}^{P_{B}}$  corresponds to the angular momentum and parity of baryon,  $J_{M}^{P_{M}}$ is for meson and  $J_{l}^{P_{l}}$ is for orbital excitation between meson and baryon. The results are listed in unit of nuclear magneton $\mu_N$.}
          \label{tab:magmom82f}
\end{table}

\begin{table}[H]
\centering
\renewcommand{\arraystretch}{1.25}
\begin{tabular}{|c|c|c|c|c|c|c|}
        \hline
    \quad    $J^P$ \quad &  \quad $^{2s+1}L_j$  \quad & \quad $J_{B}^{P_{B}} \otimes J_{M}^{P_{M}}\otimes J_{l}^{P_{l}}$ \quad &  \quad Magnetic moment \quad     \\\hline\hline
        
$\frac{1}{2}^-$  \,&\,  $^2S_{\frac{1}{2}}$ \,&\, ${\frac{1}{2}}^{+}\otimes 0^{-} \otimes 0^{+}$ \,&\, $0.022$  \\
\,&\, \,&\, ${\frac{1}{2}}^{+} \otimes 1^{-} \otimes 0^{+}$ \,&\, $0.037$\\
\,&\, \,&\, ${\frac{3}{2}}^{+} \otimes 1^{-} \otimes 0^{+}$ \,&\, $-0.059$\\ \hline

$\frac{3}{2}^-$  \,&\,  $^4S_{\frac{3}{2}}$ \,&\, ${\frac{1}{2}}^{+}\otimes 1^{+} \otimes 0^{-}$ \,&\, $0.089$  \\
 \,&\, \,&\, ${\frac{3}{2}}^{+} \otimes 0^{-} \otimes 0^{+}$ \,&\, $-0.066$  \\
 \,&\, \,&\, ${\frac{3}{2}}^{+} \otimes 1^{+} \otimes 0^{-}$ \,&\,$-0.022$ \\
\hline

$\frac{5}{2}^-$  \,&\,  $^6S_{\frac{5}{2}}$ \,&\, ${\frac{3}{2}}^{+}\otimes 1^{-} \otimes 0^{+}$ \,&\, $-0.127$  \\
\hline

$\frac{1}{2}^+$  \,&\,  $^2P_{\frac{1}{2}}$ \,&\, ${\frac{1}{2}}^{+} \otimes 0^{-} \otimes 1^{-}$ \,&\, $-0.522$  \\ 
 \,&\,  \,&\, ${(\frac{1}{2}}^{+} \otimes 1^{-})_{\frac{1}{2}} \otimes 1^{-}$ \,&\, $-0.526$  \\ 
\hline
 \,&\,  $^4P_{\frac{1}{2}}$ \,&\, $({\frac{1}{2}}^{+} \otimes 1^{-})_{\frac{3}{2}} \otimes 1^{-}$ \,&\, $0.306$  \\ 
 \,&\,   \,&\, ${\frac{3}{2}}^{+} \otimes 0^{-} \otimes 1^{-}$ \,&\, $0.220$  \\ 
\hline
 \,&\,  $^2P_{\frac{1}{2}}$ \,&\, ${(\frac{3}{2}}^{+} \otimes 1^{-})_{\frac{1}{2}} \otimes 1^{-}$ \,&\, $-0.494$  \\ \hline
 \,&\,  $^4P_{\frac{1}{2}}$ \,&\, ${(\frac{3}{2}}^{+} \otimes 1^{-})_{\frac{3}{2}} \otimes 1^{-}$ \,&\, $0.245$  \\ \hline

$\frac{3}{2}^+$  \,&\,  $^2P_{\frac{3}{2}}$ \,&\, ${\frac{1}{2}}^{+} \otimes 0^{-} \otimes 1^{-}$ \,&\, $-0.749$  \\ 
 \,&\, \,&\, ${(\frac{1}{2}}^{+} \otimes 1^{-})_{\frac{1}{2}} \otimes 1^{-}$ \,&\, $-0.734$  \\ \hline
 \,&\,  $^4P_{\frac{3}{2}}$ \,&\, ${(\frac{1}{2}}^{+} \otimes 1^{-})_{\frac{3}{2}} \otimes 1^{-}$ \,&\, $-0.243$  \\ 
\,&\, \,&\, ${\frac{3}{2}}^{+} \otimes 0^{-} \otimes 1^{-}$ \,&\, $-0.357$ \\ 
\hline
  \,&\,  $^2P_{\frac{3}{2}}$ \,&\, ${(\frac{3}{2}}^{+} \otimes 1^{-})_{\frac{1}{2}}  \otimes 1^{-}$ \,&\, $-0.830$  \\ 
\hline
  \,&\,  $^4P_{\frac{3}{2}}$ \,&\, ${(\frac{3}{2}}^{+} \otimes 1^{-})_{\frac{3}{2}}  \otimes 1^{-}$ \,&\, $-0.325$  \\ 
\hline
  \,&\,  $^6P_{\frac{3}{2}}$ \,&\, ${(\frac{3}{2}}^{+} \otimes 1^{-})_{\frac{5}{2}}  \otimes 1^{-}$ \,&\, $-0.285$  \\ 
\hline

$\frac{5}{2}^+$  \,&\,  $^4P_{\frac{5}{2}}$ \,&\, ${\frac{1}{2}}^{+} \otimes 1^{-} \otimes 1^{-}$ \,&\, $-0.683$  \\ 
 \,&\,  \,&\, ${\frac{3}{2}}^{+} \otimes 0^{-} \otimes 1^{-}$ \,&\, $-0.838$  \\ 
 \,&\,  \,&\, ${(\frac{3}{2}}^{+} \otimes 1^{-})_{\frac{3}{2}} \otimes 1^{-}$ \,&\, $-0.793$  \\
\hline
 \,&\,  $^6P_{\frac{5}{2}}$ \,&\, ${(\frac{3}{2}}^{+} \otimes 1^{-})_{\frac{5}{2}} \otimes 1^{-}$ \,&\, $-0.220$  \\

 \hline \hline
    \end{tabular}
\caption{Magnetic moments of the $10_f$ representation of the molecular hidden-bottom  pentaquark state.  $J_{B}^{P_{B}}$  corresponds to the angular momentum and parity of baryon,  $J_{M}^{P_{M}}$ is for meson and  $J_{l}^{P_{l}}$ is for orbital excitation between meson and baryon. The results are listed in unit of nuclear magneton $\mu_N$.}
\label{tab:magmom10f}
\end{table}

\section{Magnetic Moments of Hidden-Bottom Pentaquark States in Diquark-Diquark-Antiquark Model}
\label{sec:dda}
The diquark-diquark-antiquark configuration is $(b q_1) (q_2 q_3) (\bar{b})$. In this model, $P$-wave excitation includes $\rho$ mode excitation and $\lambda$ mode excitation. The $P$-wave orbital excitation of the $\rho$ mode  lies between the $(b q_1)$ diquark and $(q_2 q_3)$ diquark. The $P$-wave orbital excitation of the $\lambda$ mode lies between the antibottom quark $(\bar{b})$ and the center of mass of system of $(b q_1)$ diquark and $(q_2 q_3)$ diquark. 

The magnetic moment of the diquark-diquark-antiquark model can be written as 
\begin{eqnarray}
\hat{\mu}  = \ \hat{\mu}_{H}+\hat{\mu}_{L}+\hat{\mu}_{\bar{b}}+\hat{\mu}_{l},
\end{eqnarray}
where the subscripts $H$ and $L$ represent a heavy diquark $(b q_1)$ and light diquark $(q_2 q_3)$, respectively, and $l$ is the orbital excitation. In the diquark-diquark-antiquark model, the specific magnetic moments formula of the pentaquark $(b q_1)(q_2 q_3)\bar{b}$ is: 
\begin{eqnarray}
	\mu  &=& \langle\ \psi_{Pentaquark} \ |\ \hat{\mu}_{H}+\hat{\mu}_{L}+\hat{\mu}_{\bar{b}}+\hat{\mu}_{l}\ |\ \psi_{Pentaquark} \  \rangle\nonumber\\
	&=&\sum_{S_z,l_z}\ \langle\ SS_z,ll_z|JJ_z\ \rangle^{2}  \left \{ \mu_{l} l_z + \sum_{\widetilde{S}_{\bar{b}}}\ \langle\ S_{\bar{b}} \widetilde{S}_{\bar{b}},S_{HL} \widetilde{S}_{HL}|SS_z\ \rangle^{2} 	\Bigg [g\widetilde{S}_{\bar{b}}\mu_{\bar{b}}\nonumber\right.\\
	&+&\left.\sum_{\widetilde{S}_{H},\widetilde{S}_{L}}\ \langle\ S_{H} \widetilde{S}_{H},S_{L} \widetilde{S}_{L}|S_{HL} \widetilde{S}_{HL}\rangle^{2}\bigg(\widetilde{S}_{H}(\mu_{b}+\mu_{q_1})+\widetilde{S}_{L}(\mu_{q_2}+\mu_{q_3})\bigg )	\Bigg ]\right \}.        
\end{eqnarray}
where $S_H$ is the spin of the $(bq_1)$ diquark, $S_L$ is the spin of the $(q_2q_3)$ diquark, $S_{HL}$ is the spin of the $(bq_1)(q_2q_3)$ system and $\widetilde{S}$ is the third spin component. We use the following diquark masses from Ref. \cite{Ebert:2010af}:
\begin{eqnarray}
	[q,q]&=& 710~\text{MeV}, \ \ \ \ \{q,q\} =909~\text{MeV}, \nonumber \\\ 
	[b,q]&=& 5359~\text{MeV},\ \ \ \{b,q\} =5381~\text{MeV}, 
\end{eqnarray}
where $[\cdots]$ denotes antisymmetric and $\{\cdots\}$ denotes symmetric representation in flavour, respectively.

As mentioned before, there are orbital excitations in the $\rho$ mode and $\lambda$ mode. The corresponding magnetic moments  $\mu_{l}$ in $\rho$  mode and $\lambda$  mode can be written as
	\begin{eqnarray} 
\mu_{l,\rho}&=&\frac{{m_{q_2q_3}}\mu_{bq_1}+{m_{bq_1}}\mu_{q_2q_3}}{m_{bq_1}+m_{q_1q_2}}, \\ 
\mu_{l,\lambda}&=&\frac{{m_{\bar{b}}}\mu_{(bq_1)(q_2q_3)}+{m_{(bq_1)(q_2q_3)}}\mu_{\bar{b}}}{m_{(bq_1)(q_2q_3)}+m_{\bar{b}}},
   	\end{eqnarray}
where ${m_{(bq_1)(q_2q_3)}}=\frac{{m_{q_2q_3}}{m_{bq_1}}}{m_{bq_1}+m_{q_2q_3}}$, $ m $ and $ \mu $ represent the mass and magnetic moment of the corresponding clusters represented by their subscripts: $m_{bq_1}$ is the mass of $bq_1$ diquark and $m_{q_2q_3}$ is the mass of $q_2q_3$ diquark. Since the $S$-wave systems are more accessible in the experiments, especially for $b$ quarks, we obtain magnetic moments of $S$-wave diquark-diquark-antiquark states. The results are listed in Table \ref{tab:magmomsdda}. Excited $P$-wave magnetic moments are given in Table \ref{tab:magmomsddaexcited81f} for $8_{1f}$ representation, Table \ref{tab:magmomsddaexcited82f} for $8_{2f}$ representation, and Table \ref{tab:magmomsddaexcited10f} for $10_{f}$ representation, respectively.

\begin{table}[H]
    \centering
    \renewcommand{\arraystretch}{1.25}
    \begin{tabular}{|c|c|c|c|c|c|c|}
        \hline
     \quad   $J^P$ \quad &\quad  $^{2s+1}L_j$ \quad & \quad $J_{H}^{P_H} \otimes J_{L}^{P_L}\otimes J_{\bar{b}}^{P_{\bar{b}}} \otimes J_{l}^{P_{l}}$ \quad & \quad  Magnetic moment  \quad \\\hline\hline
        
        &    & $8_{1f}$ &    \\\hline\hline  

$\frac{1}{2}^-$  \,&\,  $^2S_{\frac{1}{2}}$ \,&\, $0^{+}\otimes1^{+}\otimes{\frac{1}{2}}^{-}\otimes0^{+}$ \,&\, $-1.928$  \\ 

 \,&\,  \,&\, $(1^{+}\otimes1^{+})_{0} \otimes {\frac{1}{2}}^{-}\otimes0^{+}$ \,&\,  $0.066$   \\ 
 
 \,&\,   \,&\, $(1^{+}\otimes1^{+})_{1} \otimes {\frac{1}{2}}^{-}\otimes0^{+}$ \,&\, $-0.044$  \\ \hline

$\frac{3}{2}^-$  \,&\, $^4S_{\frac{3}{2}}$ \,&\, $(0^{+}\otimes1^{+}) \otimes {\frac{1}{2}}^{-}\otimes0^{+}$ \,&\, $-0.864$  \\ 

 \,&\,  \,&\, $(1^{+}\otimes1^{+})_{1} \otimes {\frac{1}{2}}^{-}\otimes0^{+}$ \,&\, $0.033$  \\ 

 \,&\,  \,&\, $(1^{+}\otimes1^{+})_{2} \otimes {\frac{1}{2}}^{-}\otimes0^{+}$ \,&\, $-0.099$  \\ \hline

$\frac{5}{2}^-$  \,&\,  $^6S_{\frac{5}{2}}$ \,&\, $1^{+}\otimes1^{+}\otimes{\frac{1}{2}}^{-}\otimes0^{+}$ \,&\, $2.792$  \\ \hline \hline

 &    & $8_{2f}$ &    \\\hline\hline        
        
$\frac{1}{2}^-$  \,&\,  $^2S_{\frac{1}{2}}$ \,&\,  $0^{+}\otimes0^{+}\otimes{\frac{1}{2}}^{-}\otimes0^{+}$ \,&\, $0.066$  \\

\,&\, \,&\, $1^{+}\otimes0^{+}\otimes{\frac{1}{2}}^{-}\otimes0^{+}$ \,&\, $-0.687$\\ \hline

$\frac{3}{2}^-$  \,&\,  $^4S_{\frac{3}{2}}$ \,&\, $1^{+}\otimes0^{+}\otimes{\frac{1}{2}}^{-}\otimes0^{+}$ \,&\, $-0.930$  \\
\hline \hline

&    & $10_{f}$ &    \\\hline\hline

$\frac{1}{2}^-$  \,&\,  $^2S_{\frac{1}{2}}$ \,&\, $0^{+}\otimes1^{+}\otimes{\frac{1}{2}}^{-}\otimes0^{+}$ \,&\, $-0.066$  \\ 

  \,&\,   \,&\, $(1^{+}\otimes1^{+})_{0} \otimes {\frac{1}{2}}^{-}\otimes0^{+}$ \,&\, $0.066$  \\ 
  
  \,&\,   \,&\, $(1^{+}\otimes1^{+})_{0} \otimes {\frac{1}{2}}^{-}\otimes0^{+}$ \,&\, $-0.044$  \\  \hline
  
$\frac{3}{2}^-$  \,&\,  $^4S_{\frac{3}{2}}$ \,&\, $(0^{+}\otimes1^{+}) \otimes {\frac{1}{2}}^{-}\otimes0^{+}$ \,&\, $0.066$  \\

 \,&\,   \,&\, $(1^{+}\otimes1^{+})_{1} \otimes {\frac{1}{2}}^{-}\otimes0^{+}$ \,&\, $0.033$  \\   
  
 \,&\,   \,&\, $(1^{+}\otimes1^{+})_{2} \otimes {\frac{1}{2}}^{-}\otimes0^{+}$ \,&\, $-0.099$  \\     \hline
  
$\frac{5}{2}^-$  \,&\,  $^6S_{\frac{3}{2}}$ \,&\, $1^{+}\otimes1^{+}\otimes{\frac{1}{2}}^{-}\otimes0^{+}$ \,&\, $2.792$  \\ 
 \hline \hline
    \end{tabular}
        \caption{Magnetic moments of the $S$-wave diquark-diquark-antiquark hidden-bottom pentaquark states in $8_{1f}$, $8_{2f}$ and $10_f$ representations. $J_{H}^{P_H} $ corresponds to the angular momentum and parity of $(bq_1)$, $J_{L}^{P_L}$ is for $(q_2q_3)$, $J_{\bar{b}}^{P_{\bar{b}}}$ is for $\bar{b}$, and $J_{l}^{P_{l}}$ is for orbital. The results are listed in unit of nuclear magneton $\mu_N$.}
\label{tab:magmomsdda}
\end{table}

\begin{table}[H]
    \centering
    \renewcommand{\arraystretch}{1.25}
    \begin{tabular}{|c|c|c|c|c|c|c|}
        \hline
     \quad   $J^P$ \quad &\quad  $^{2s+1}L_j$ \quad & \quad $J_{H}^{P_H} \otimes J_{L}^{P_L}\otimes J_{\bar{b}}^{P_{\bar{b}}} \otimes J_{l}^{P_{l}}$ \quad & \quad  Magnetic moment $(\lambda)$ \quad & \quad  Magnetic moment $(\rho)$ \quad \\\hline\hline

$\frac{1}{2}^{+}$  \,&\,  ${^{2}P_{\frac{1}{2}}}$ \,&\, $(0^{+}\otimes1^{+}\otimes {\frac{1}{2}}^{-})_{\frac{1}{2}}\otimes1^{-}$  \,&\, $0.347$  \,&\, $0.104$ \\ 

 \,&\,  \,&\, $((1^{+}\otimes1^{+})_{0}\otimes{\frac{1}{2}}^{-})_{\frac{1}{2}}\otimes1^{-}$ \,&\,  $0.108$   \,&\, $-0.043$ \\ 
 
 \,&\,   \,&\, $((1^{+}\otimes1^{+})_{1}\otimes{\frac{1}{2}}^{-})_{\frac{1}{2}}\otimes1^{-}$ \,&\, $0.145$ \,&\, $-0.006$ \\ \hline

 \,&\, ${^{4}P_{\frac{1}{2}}}$ \,&\, $(0^{+}\otimes1^{+}\otimes {\frac{1}{2}}^{-})_{\frac{3}{2}}\otimes1^{-}$ \,&\, $-0.371$ \,&\, $-0.331$\\ 

 \,&\,  \,&\, $((1^{+}\otimes1^{+})_{1}\otimes{\frac{1}{2}}^{-})_{\frac{3}{2}}\otimes1^{-}$ \,&\, $-0.130$ \,&\, $-0.090$\\ 
 
 \,&\,  \,&\, $((1^{+}\otimes1^{+})_{2}\otimes{\frac{1}{2}}^{-})_{\frac{3}{2}}\otimes1^{-}$ \,&\, $-0.204$ \,&\, $-0.189$\\ \hline

$\frac{3}{2}^{+}$  \,&\,  ${^{2}P_{\frac{3}{2}}}$ \,&\, $(0^{+}\otimes1^{+}\otimes {\frac{1}{2}}^{-})_{\frac{1}{2}}\otimes1^{-}$  \,&\, $-0.332$  \,&\, $-0.452$ \\ 

  \,&\,   \,&\, $((1^{+}\otimes1^{+})_{0}\otimes{\frac{1}{2}}^{-})_{\frac{1}{2}}\otimes1^{-}$  \,&\, $0.139$  \,&\, $0.018$ \\ 

  \,&\,   \,&\, $((1^{+}\otimes1^{+})_{1}\otimes{\frac{1}{2}}^{-})_{\frac{1}{2}}\otimes1^{-}$  \,&\, $0.080$  \,&\, $-0.040$ \\ \hline

  \,&\,  ${^{4}P_{\frac{3}{2}}}$ \,&\, $(0^{+}\otimes1^{+}\otimes {\frac{1}{2}}^{-})_{\frac{3}{2}}\otimes1^{-}$  \,&\, $-0.553$  \,&\, $-0.644$ \\ 

  \,&\,   \,&\, $((1^{+}\otimes1^{+})_{1}\otimes{\frac{1}{2}}^{-})_{\frac{3}{2}}\otimes1^{-}$ \,&\, $0.102$  \,&\, $0.011$ \\ 

  \,&\,   \,&\, $((1^{+}\otimes1^{+})_{2}\otimes{\frac{1}{2}}^{-})_{\frac{3}{2}}\otimes1^{-}$ \,&\, $0.04$  \,&\, $-0.086$ \\ \hline

  \,&\,  ${^{6}P_{\frac{3}{2}}}$ \,&\,$((1^{+}\otimes1^{+})_{2}\otimes{\frac{1}{2}}^{-})_{\frac{5}{2}}\otimes1^{-}$  \,&\, $-0.462$  \,&\, $-0.327$ \\ \hline

$\frac{5}{2}^{+}$  \,&\,  ${^{4}P_{\frac{5}{2}}}$ \,&\, $(0^{+}\otimes1^{+}\otimes {\frac{1}{2}}^{-})\otimes1^{-}$  \,&\, $-0.664$  \,&\, $-0.890$ \\ 

 \,&\,   \,&\,$((1^{+}\otimes1^{+})_{1}\otimes{\frac{1}{2}}^{-})_{\frac{3}{2}}\otimes1^{-}$ \,&\, $0.227$  \,&\, $0.001$ \\

 \,&\,   \,&\,$((1^{+}\otimes1^{+})_{2}\otimes{\frac{1}{2}}^{-})_{\frac{3}{2}}\otimes1^{-}$ \,&\, $0.157$  \,&\, $0.036$ \\  \hline

 \,&\,  ${^{6}P_{\frac{5}{2}}}$  \,&\, $((1^{+}\otimes1^{+})_{2}\otimes{\frac{1}{2}}^{-})_{\frac{5}{2}}\otimes1^{-}$  \,&\, $0.055$  \,&\, $-0.008$ \\ 

\hline \hline
    \end{tabular}
        \caption{Magnetic moments of the excited diquark-diquark-antiquark hidden-bottom pentaquark states in $8_{1f}$ representation. $J_{H}^{P_H} $ corresponds to the angular momentum and parity of $(bq_1)$, $J_{L}^{P_L}$ is for $(q_2q_3)$, $J_{\bar{b}}^{P_{\bar{b}}}$ is for $\bar{b}$, and $J_{l}^{P_{l}}$ is for orbital. The results are listed in unit of nuclear magneton $\mu_N$.}
\label{tab:magmomsddaexcited81f}
\end{table}

\begin{table}[H]
    \centering
    \renewcommand{\arraystretch}{1.25}
    \begin{tabular}{|c|c|c|c|c|c|c|}
        \hline
     \quad   $J^P$ \quad &\quad  $^{2s+1}L_j$ \quad & \quad $J_{H}^{P_H} \otimes J_{L}^{P_L}\otimes J_{\bar{b}}^{P_{\bar{b}}} \otimes J_{l}^{P_{l}}$ \quad & \quad  Magnetic moment $(\lambda)$ \quad & \quad  Magnetic moment $(\rho)$ \quad \\\hline\hline

$\frac{1}{2}^{+}$  \,&\,  ${^{2}P_{\frac{1}{2}}}$ \,&\, $0^{+}\otimes0^{+}\otimes{\frac{1}{2}}^{-}\otimes1^{-}$  \,&\, $0.154$  \,&\, $0.236$ \\ 

 \,&\,  \,&\, $(1^{+}\otimes0^{+}\otimes{\frac{1}{2}}^{-})_{\frac{1}{2}}\otimes1^{-}$ \,&\,  $0.401$   \,&\, $0.309$ \\

 \,&\, ${^{4}P_{\frac{1}{2}}}$ \,&\, $(1^{+}\otimes0^{+}\otimes{\frac{1}{2}}^{-})_{\frac{3}{2}}\otimes1^{-}$ \,&\, $-1.437$ \,&\, $-0.556$\\ \hline

$\frac{3}{2}^{+}$  \,&\,  ${^{2}P_{\frac{3}{2}}}$ \,&\, $0^{+}\otimes0^{+}\otimes{\frac{1}{2}}^{-}\otimes1^{-}$  \,&\, $0.331$  \,&\, $0.186$ \\

 \,&\,  \,&\, $(1^{+}\otimes0^{+}\otimes{\frac{1}{2}}^{-})_{\frac{1}{2}}\otimes1^{-}$  \,&\, $-0.429$  \,&\, $-0.566$ \\ \hline
 
 \,&\, ${^{4}P_{\frac{3}{2}}}$ \,&\, $(1^{+}\otimes0^{+}\otimes{\frac{1}{2}}^{-})_{\frac{3}{2}}\otimes1^{-}$  \,&\, $-0.579$  \,&\, $-0.634$ \\ \hline 
 
$\frac{5}{2}^{+}$ \,&\, ${^{4}P_{\frac{5}{2}}}$ \,&\, $1^{+}\otimes0^{+}\otimes{\frac{1}{2}}^{-}\otimes1^{-}$  \,&\, $-0.672$  \,&\, $-0.810$ \\

\hline \hline
    \end{tabular}
        \caption{Magnetic moments of the excited diquark-diquark-antiquark hidden-bottom pentaquark states in $8_{2f}$ representation. $J_{H}^{P_H} $ corresponds to the angular momentum and parity of $(bq_1)$, $J_{L}^{P_L}$ is for $(q_2q_3)$, $J_{\bar{b}}^{P_{\bar{b}}}$ is for $\bar{b}$, and $J_{l}^{P_{l}}$ is for orbital. The results are listed in unit of nuclear magneton $\mu_N$.}
\label{tab:magmomsddaexcited82f}
\end{table}

\begin{table}[H]
    \centering
    \renewcommand{\arraystretch}{1.25}
    \begin{tabular}{|c|c|c|c|c|c|c|}
        \hline
     \quad   $J^P$ \quad &\quad  $^{2s+1}L_j$ \quad & \quad $J_{H}^{P_H} \otimes J_{L}^{P_L}\otimes J_{\bar{b}}^{P_{\bar{b}}} \otimes J_{l}^{P_{l}}$ \quad & \quad  Magnetic moment $(\lambda)$ \quad & \quad  Magnetic moment $(\rho)$ \quad \\\hline\hline

$\frac{1}{2}^{+}$  \,&\,  ${^{2}P_{\frac{1}{2}}}$ \,&\, $(0^{+}\otimes1^{+}\otimes {\frac{1}{2}}^{-})_{\frac{1}{2}}\otimes1^{-}$  \,&\, $0.141$  \,&\, $0.030$ \\ 

 \,&\,  \,&\, $((1^{+}\otimes1^{+})_{0}\otimes{\frac{1}{2}}^{-})_{\frac{1}{2}}\otimes1^{-}$ \,&\,  $0.108$   \,&\, $-0.001$ \\ 
 
 \,&\,   \,&\, $((1^{+}\otimes1^{+})_{1}\otimes{\frac{1}{2}}^{-})_{\frac{1}{2}}\otimes1^{-}$ \,&\, $0.145$ \,&\, $0.035$ \\ \hline

 \,&\, ${^{4}P_{\frac{1}{2}}}$ \,&\, $(0^{+}\otimes1^{+}\otimes {\frac{1}{2}}^{-})_{\frac{3}{2}}\otimes1^{-}$ \,&\, $-0.030$ \,&\, $0.025$\\ 

 \,&\,  \,&\, $((1^{+}\otimes1^{+})_{1}\otimes{\frac{1}{2}}^{-})_{\frac{3}{2}}\otimes1^{-}$ \,&\, $-0.046$ \,&\, $0.007$\\ 
 
 \,&\,  \,&\, $((1^{+}\otimes1^{+})_{2}\otimes{\frac{1}{2}}^{-})_{\frac{3}{2}}\otimes1^{-}$ \,&\, $0.292$ \,&\, $0.346$\\ \hline

$\frac{3}{2}^{+}$  \,&\,  ${^{2}P_{\frac{3}{2}}}$ \,&\, $(0^{+}\otimes1^{+}\otimes {\frac{1}{2}}^{-})_{\frac{1}{2}}\otimes1^{-}$  \,&\, $0.178$  \,&\, $0.012$ \\ 

  \,&\,   \,&\, $((1^{+}\otimes1^{+})_{0}\otimes{\frac{1}{2}}^{-})_{\frac{1}{2}}\otimes1^{-}$  \,&\, $0.260$  \,&\, $0.097$ \\ 

  \,&\,   \,&\, $((1^{+}\otimes1^{+})_{1}\otimes{\frac{1}{2}}^{-})_{\frac{1}{2}}\otimes1^{-}$  \,&\, $0.150$  \,&\, $-0.013$ \\ \hline

  \,&\,  ${^{4}P_{\frac{3}{2}}}$ \,&\, $(0^{+}\otimes1^{+}\otimes {\frac{1}{2}}^{-})_{\frac{3}{2}}\otimes1^{-}$  \,&\, $0.129$  \,&\, $0.062$ \\ 

  \,&\,   \,&\, $((1^{+}\otimes1^{+})_{1}\otimes{\frac{1}{2}}^{-})_{\frac{3}{2}}\otimes1^{-}$ \,&\, $0.102$  \,&\, $0.176$ \\ 

  \,&\,   \,&\, $((1^{+}\otimes1^{+})_{2}\otimes{\frac{1}{2}}^{-})_{\frac{3}{2}}\otimes1^{-}$ \,&\, $0.004$  \,&\, $-0.060$ \\ \hline

  \,&\,  ${^{6}P_{\frac{3}{2}}}$ \,&\,$((1^{+}\otimes1^{+})_{2}\otimes{\frac{1}{2}}^{-})_{\frac{5}{2}}\otimes1^{-}$  \,&\, $-0.462$  \,&\, $-0.365$ \\ \hline

$\frac{5}{2}^{+}$  \,&\,  ${^{4}P_{\frac{5}{2}}}$ \,&\, $(0^{+}\otimes1^{+}\otimes {\frac{1}{2}}^{-})\otimes1^{-}$  \,&\, $0.267$  \,&\, $0.100$ \\ 

 \,&\,   \,&\,$((1^{+}\otimes1^{+})_{1}\otimes{\frac{1}{2}}^{-})_{\frac{3}{2}}\otimes1^{-}$ \,&\, $0.227$  \,&\, $0.064$ \\

 \,&\,   \,&\,$((1^{+}\otimes1^{+})_{2}\otimes{\frac{1}{2}}^{-})_{\frac{3}{2}}\otimes1^{-}$ \,&\, $0.294$  \,&\, $0.131$ \\  \hline

 \,&\,  ${^{6}P_{\frac{5}{2}}}$  \,&\, $((1^{+}\otimes1^{+})_{2}\otimes{\frac{1}{2}}^{-})_{\frac{5}{2}}\otimes1^{-}$  \,&\, $0.055$  \,&\, $0.006$ \\ 

\hline \hline
    \end{tabular}
        \caption{Magnetic moments of the excited diquark-diquark-antiquark hidden-bottom pentaquark states in $10_{f}$ representation. $J_{H}^{P_H} $ corresponds to the angular momentum and parity of $(bq_1)$, $J_{L}^{P_L}$ is for $(q_2q_3)$, $J_{\bar{b}}^{P_{\bar{b}}}$ is for $\bar{b}$, and $J_{l}^{P_{l}}$ is for orbital. The results are listed in unit of nuclear magneton $\mu_N$.}
\label{tab:magmomsddaexcited10f}
\end{table}

\section{Magnetic Moments of Hidden-Bottom Pentaquark States in Diquark-Triquark Model}
\label{sec:ditri}
The diquark-triquark configuration is $(bq_1)(\bar{b}q_2q_3)$. The total magnetic moment in diquark-triquark model is 
\begin{eqnarray}
	\hat{\mu}  = \ \hat{\mu}_{D}+\hat{\mu}_{T}+\hat{\mu}_{l}, 
\end{eqnarray}
where the $D$ denotes diquark, $T$ denotes triquark, and $l$ is the orbital excitation between the diquark and triquark. The magnetic moment formula can be written as
\begin{eqnarray}
	\mu  
	&=& \langle\ \psi_{Pentaquark}\ |\ \hat{\mu}_{D}+\hat{\mu}_{T}+\hat{\mu}_{l}\ |\ \psi_{Pentaquark}\  \rangle\nonumber\\
	&=&
	\sum_{S_z,l_z}\ \langle\ SS_z,ll_z|JJ_z\ \rangle^{2}  \left \{ \mu_{l} l_z 	+
	\sum_{\widetilde{S}_{\mathcal{D}},\widetilde{S}_{D}}\ \langle\ S_\mathcal{D} \widetilde{S}_{\mathcal{D}},S_T \widetilde{S}_{T}|SS_z\ \rangle^{2} \Bigg [
	\widetilde{S}_{D}\bigg(\mu_{b} + \mu_{q_1}\bigg )\nonumber\right.\\
	&+&\left.
	\sum_{\widetilde{S}_{\bar{b}}}\ \langle\ S_{\bar{b}} \widetilde{S}_{\bar{b}},S_{m} \widetilde{S}_{T}-\widetilde{S}_{\bar{b}}|S_{T} \widetilde{S}_{T}\rangle^{2}\bigg(g\mu_{\bar{b}}\widetilde{S}_{\bar{b}}+(\widetilde{S}_{T}-\widetilde{S}_{\bar{b}})(\mu_{q_2}+\mu_{q_3})\bigg )
	\Bigg ]\right \},          
\end{eqnarray}
where $S_D$, $S_T$ and $S_m$ represent the  diquark, triquark and the light diquark spin inside the triquark, respectively.  The orbital magnetic moment in diquark-triquark model can be written as
	\begin{eqnarray} 
		{\mu_l}&=&\frac{{m_{bq_{1}}}\mu_{\bar{b}q_{2}q_{3} }+{m_{\bar{b}q_{2}q_{2} }}\mu_{bq_{1}}}{m_{\bar{b}q_{2}q_{3}}+m_{bq_{1}}},
	\end{eqnarray}
where $m$ and $\mu$ are the mass and magnetic moment of the cluster represented by their subscripts. The mass of the triquark is approximately equal to the sum of the corresponding diquark mass and antiquark mass \cite{Wang:2016dzu}. The results for the magnetic moments of diquark-triquark model are listed in Table \ref{tab:magmomsdtm81f} for $8_{1f}$, Table \ref{tab:magmomsdtm82f} for $8_{2f}$, and Table \ref{tab:magmomdtm10f} for $10_{f}$ representations, respectively.

\begin{table}[H]
    \centering
    \renewcommand{\arraystretch}{1.25}
    \begin{tabular}{|c|c|c|c|c|c|c|}
        \hline
     \quad   $J^P$ \quad &\quad  $^{2s+1}L_j$ \quad & \quad $J_{T}^{P_T} \otimes J_{D}^{P_D} \otimes J_{l}^{P_{l}}$ \quad & \quad  Magnetic moment  \quad   \\\hline\hline

$\frac{1}{2}^{-}$  \,&\,  ${^{2}S_{\frac{1}{2}}}$ \,&\, ${\frac{1}{2}}^{-}\otimes0^{+}\otimes0^{+}$  \,&\, $-0.642$   \\ 

 \,&\,  \,&\, ${\frac{1}{2}}^{-}\otimes1^{+} \otimes 0^{+}$ \,&\,  $0.790$   \\

 \,&\,  \,&\, ${\frac{3}{2}}^{-}\otimes1^{+} \otimes 0^{+}$ \,&\,  $-0.768$   \\ \hline

$\frac{3}{2}^{-}$ \,&\, ${^{4}S_{\frac{3}{2}}}$ \,&\, ${\frac{1}{2}}^{-}\otimes1^{+}\otimes0^{-}$ \,&\, $0.221$ \\ 

 \,&\,  \,&\, ${\frac{3}{2}}^{-}\otimes0^{+}\otimes0^{+}$ \,&\, $-0.864$ \\ 

 \,&\,  \,&\, ${\frac{3}{2}}^{-}\otimes1^{+}\otimes0^{-}$ \,&\, $-0.288$ \\ \hline

$\frac{5}{2}^{-}$  \,&\,  ${^{6}S_{\frac{5}{2}}}$ \,&\, ${\frac{3}{2}}^{-}\otimes1^{+}\otimes0^{+}$  \,&\, $-1.231$   \\ \hline

$\frac{1}{2}^{+}$ \,&\, ${^{2}P_{\frac{1}{2}}}$ \,&\, ${\frac{1}{2}}^{-}\otimes0^{+}\otimes1^{-}$ \,&\, $0.237$   \\ 
 
 \,&\,  \,&\, $({\frac{1}{2}}^{-}\otimes1^{+})_{\frac{1}{2}}\otimes1^{-}$ \,&\, $-0.257$   \\ \hline

 \,&\, ${^{4}P_{\frac{1}{2}}}$ \,&\, $({\frac{1}{2}}^{-}\otimes1^{+})_{\frac{3}{2}}\otimes1^{-}$ \,&\, $0.061$   \\

 \,&\,  \,&\, ${\frac{3}{2}}^{-}\otimes0^{+}\otimes1^{-}$  \,&\, $-0.491$ \\  \hline
 
 \,&\, ${^{2}P_{\frac{1}{2}}}$ \,&\, $({\frac{3}{2}}^{-}\otimes1^{+})_{\frac{1}{2}}\otimes1^{-}$ \,&\, $0.249$   \\  \hline
 
  \,&\, ${^{4}P_{\frac{1}{2}}}$ \,&\, $({\frac{3}{2}}^{-}\otimes1^{+})_{\frac{3}{2}}\otimes1^{-}$ \,&\, $-0.157$   \\  \hline
 
$\frac{3}{2}^{+}$  \,&\,  ${^{2}P_{\frac{3}{2}}}$ \,&\, ${\frac{1}{2}}^{-}\otimes0^{+}\otimes1^{-}$  \,&\, $-0.609$   \\ 

  \,&\,   \,&\, $({\frac{1}{2}}^{-}\otimes1^{+})_{\frac{1}{2}}\otimes1^{-}$  \,&\, $0.779$   \\ \hline
 
  \,&\,  ${^{4}P_{\frac{3}{2}}}$ \,&\, $({\frac{1}{2}}^{-}\otimes1^{+})_{\frac{3}{2}}\otimes1^{-}$  \,&\, $0.158$   \\ 

  \,&\,  \,&\, ${\frac{3}{2}}^{-}\otimes0^{+}\otimes1^{-}$ \,&\, $-0.620$ \\ \hline

  \,&\,  ${^{2}P_{\frac{3}{2}}}$ \,&\, $({\frac{3}{2}}^{-}\otimes1^{+})_{\frac{1}{2}}\otimes1^{-}$  \,&\, $-0.779$   \\ \hline
  
   \,&\,  ${^{4}P_{\frac{3}{2}}}$ \,&\, $({\frac{3}{2}}^{-}\otimes1^{+})_{\frac{3}{2}}\otimes1^{-}$  \,&\, $-0.215$   \\ \hline 

   \,&\, ${^{6}P_{\frac{3}{2}}}$ \,&\, $({\frac{3}{2}}^{-}\otimes1^{+})_{\frac{5}{2}}\otimes1^{-}$  \,&\, $0.006$   \\ \hline 

$\frac{5}{2}^{+}$  \,&\,${^{4}P_{\frac{5}{2}}}$ \,&\, ${\frac{1}{2}}^{-}\otimes1^{+}\otimes1^{-}$  \,&\, $0.211$   \\ 

 \,&\, \,&\, ${\frac{3}{2}}^{-}\otimes0^{+}\otimes1^{-}$  \,&\, $-0.830$   \\ 

 \,&\, \,&\, $({\frac{3}{2}}^{-}\otimes1^{+})_{\frac{3}{2}}\otimes1^{-}$  \,&\, $-0.298$   \\ \hline

  \,&\,${^{6}P_{\frac{5}{2}}}$ \,&\, $({\frac{3}{2}}^{-}\otimes1^{+})_{\frac{5}{2}}\otimes1^{-}$  \,&\, $-0.002$   \\ 

\hline \hline
    \end{tabular}
\caption{Magnetic moments of the diquark-triquark hidden-bottom pentaquark states in $8_{1f}$ representation. $J_{T}^{P_T} $ corresponds to the angular momentum and parity of triquark, $J_{D}^{P_D}$ is for diquark, $J_{\bar{b}}$  is for anti-bottom, and $J_{l}^{P_{l}}$ is for orbital. The results are listed in unit of nuclear magneton $\mu_N$.}
\label{tab:magmomsdtm81f}
\end{table}

\begin{table}[H]
    \centering
    \renewcommand{\arraystretch}{1.25}
    \begin{tabular}{|c|c|c|c|c|c|c|}
        \hline
     \quad   $J^P$ \quad &\quad  $^{2s+1}L_j$ \quad & \quad $J_{T}^{P_T} \otimes J_{D}^{P_D} \otimes J_{l}^{P_{l}}$ \quad & \quad  Magnetic moment  \quad   \\\hline\hline

$\frac{1}{2}^{-}$  \,&\,  ${^{2}S_{\frac{1}{2}}}$ \,&\, ${\frac{1}{2}}^{-}\otimes0^{+}\otimes0^{+}$  \,&\, $0.066$   \\ 

 \,&\,  \,&\, ${\frac{1}{2}}^{-}\otimes1^{+} \otimes 0^{+}$ \,&\,  $-0.366$   \\  \hline

$\frac{3}{2}^{-}$  \,&\,  ${^{4}S_{\frac{3}{2}}}$ \,&\, ${\frac{1}{2}}^{-}\otimes1^{+}\otimes0^{+}$  \,&\, $-0.930$   \\ \hline

$\frac{1}{2}^{+}$  \,&\,  ${^{2}P_{\frac{1}{2}}}$ \,&\, ${\frac{1}{2}}^{-}\otimes0^{+}\otimes1^{-}$  \,&\, $-0.017$   \\ 

  \,&\,  \,&\, $({\frac{1}{2}}^{-}\otimes1^{+})_{\frac{1}{2}}\otimes1^{-}$ \,&\, $0.232$   \\  \hline
  
 \,&\,  ${^{4}P_{\frac{1}{2}}}$ \,&\, $({\frac{1}{2}}^{-}\otimes1^{+})_{\frac{3}{2}}\otimes1^{-}$  \,&\, $-0.519$   \\ \hline

$\frac{3}{2}^{+}$  \,&\, ${^{2}P_{\frac{3}{2}}}$ \,&\, ${\frac{1}{2}}^{-}\otimes0^{+}\otimes1^{-}$  \,&\, $0.074$   \\ 

 \,&\,  \,&\, $({\frac{1}{2}}^{-}\otimes1^{+})_{\frac{1}{2}}\otimes1^{-}$  \,&\, $-0.682$   \\ \hline
 
  \,&\, ${^{4}P_{\frac{3}{2}}}$ \,&\, $({\frac{1}{2}}^{-}\otimes1^{+})_{\frac{3}{2}}\otimes1^{-}$  \,&\, $-0.226$   \\  \hline

$\frac{5}{2}^{+}$ \,&\, ${^{4}P_{\frac{5}{2}}}$ \,&\, ${\frac{1}{2}}^{-}\otimes1^{+}\otimes1^{-}$  \,&\, $-0.925$   \\ 
\hline \hline
    \end{tabular}
\caption{Magnetic moments of the diquark-triquark hidden-bottom pentaquark states in $8_{2f}$ representation.  $J_{T}^{P_T} $ corresponds to the angular momentum and parity of triquark, $J_{D}^{P_D}$ is for diquark, $J_{\bar{b}}$  is for anti-bottom, and $J_{l}^{P_{l}}$ is for orbital. The results are listed in unit of nuclear magneton $\mu_N$.}
\label{tab:magmomsdtm82f}
\end{table}


\begin{table}[H]
\centering
\renewcommand{\arraystretch}{1.25}
\begin{tabular}{|c|c|c|c|c|c|c|}
        \hline
   \quad   $J^P$ \quad &\quad  $^{2s+1}L_j$ \quad & \quad $J_{T}^{P_T} \otimes J_{D}^{P_D} \otimes J_{l}^{P_{l}}$ \quad & \quad  Magnetic moment  \quad     \\\hline\hline
        
$\frac{1}{2}^-$  \,&\,  $^2S_{\frac{1}{2}}$ \,&\, ${\frac{1}{2}}^{-}\otimes0^{+}\otimes0^{+}$ \,&\, $-0.022$  \\
\,&\, \,&\, ${\frac{1}{2}}^{-}\otimes1^{+}\otimes0^{+}$ \,&\, $-0.037$\\
\,&\, \,&\, ${\frac{3}{2}}^{-}\otimes1^{+}\otimes0^{+}$ \,&\, $0.059$\\ \hline

$\frac{3}{2}^-$  \,&\,  $^4S_{\frac{3}{2}}$ \,&\, ${\frac{1}{2}}^{-}\otimes1^{+}\otimes0^{-}$ \,&\, $-0.089$  \\
 \,&\, \,&\, ${\frac{3}{2}}^{-}\otimes0^{+}\otimes0^{+}$ \,&\, $0.066$  \\
 \,&\, \,&\, ${\frac{3}{2}}^{-}\otimes1^{+}\otimes0^{-}$ \,&\,$0.022$ \\
\hline

$\frac{5}{2}^-$  \,&\,  $^6S_{\frac{5}{2}}$ \,&\,  ${\frac{3}{2}}^{-}\otimes1^{+}\otimes0^{+}$ \,&\, $0.127$  \\
\hline

$\frac{1}{2}^+$  \,&\,  $^2P_{\frac{1}{2}}$ \,&\, ${\frac{1}{2}}^{-}\otimes0^{+}\otimes1^{-}$ \,&\, $0.467$  \\ 
 \,&\,  \,&\,$({\frac{1}{2}}^{-}\otimes1^{+})_{\frac{1}{2}}\otimes1^{-}$ \,&\, $0.625$  \\ 
\hline
 \,&\,  $^4P_{\frac{1}{2}}$ \,&\, $({\frac{1}{2}}^{-}\otimes1^{+})_{\frac{3}{2}}\otimes1^{-}$ \,&\, $-0.286$  \\ 
 \,&\,   \,&\,  ${\frac{3}{2}}^{-}\otimes0^{+}\otimes1^{-}$  \,&\, $-0.158$  \\ 
\hline
 \,&\,  $^2P_{\frac{1}{2}}$ \,&\, $({\frac{3}{2}}^{-}\otimes1^{+})_{\frac{1}{2}}\otimes1^{-}$ \,&\, $0.567$  \\ \hline
 \,&\,  $^4P_{\frac{1}{2}}$ \,&\, $({\frac{3}{2}}^{-}\otimes1^{+})_{\frac{3}{2}}\otimes1^{-}$ \,&\, $-0.199$  \\ \hline

$\frac{3}{2}^+$  \,&\,  $^2P_{\frac{3}{2}}$ \,&\, ${\frac{1}{2}}^{-}\otimes0^{+}\otimes1^{-}$ \,&\, $0.834$  \\ 
 \,&\, \,&\, $({\frac{1}{2}}^{-}\otimes1^{+})_{\frac{1}{2}}\otimes1^{-}$ \,&\, $0.684$  \\ \hline
 \,&\,  $^4P_{\frac{3}{2}}$ \,&\, $({\frac{1}{2}}^{-}\otimes1^{+})_{\frac{3}{2}}\otimes1^{-}$ \,&\, $0.268$  \\ 
\,&\, \,&\, ${\frac{3}{2}}^{-}\otimes0^{+}\otimes1^{-}$ \,&\, $0.334$ \\ 
\hline
  \,&\,  $^2P_{\frac{3}{2}}$ \,&\, $({\frac{3}{2}}^{-}\otimes1^{+})_{\frac{1}{2}}\otimes1^{-}$ \,&\, $0.902$  \\ 
\hline
  \,&\,  $^4P_{\frac{3}{2}}$ \,&\, $({\frac{3}{2}}^{-}\otimes1^{+})_{\frac{3}{2}}\otimes1^{-}$ \,&\, $0.396$  \\ 
\hline
  \,&\,  $^6P_{\frac{3}{2}}$ \,&\, $({\frac{3}{2}}^{-}\otimes1^{+})_{\frac{5}{2}}\otimes1^{-}$ \,&\, $0.246$  \\ 
\hline

$\frac{5}{2}^+$  \,&\,  $^4P_{\frac{5}{2}}$ \,&\,${\frac{1}{2}}^{-}\otimes1^{+}\otimes1^{-}$ \,&\, $0.692$  \\ 
 \,&\,  \,&\, ${\frac{3}{2}}^{-}\otimes0^{+}\otimes1^{-}$ \,&\, $0.823$  \\ 
 \,&\,  \,&\, $({\frac{3}{2}}^{-}\otimes1^{+})_{\frac{3}{2}}\otimes1^{-}$ \,&\, $0.751$  \\
\hline
 \,&\,  $^6P_{\frac{5}{2}}$ \,&\, $({\frac{3}{2}}^{-}\otimes1^{+})_{\frac{5}{2}}\otimes1^{-}$ \,&\, $0.233$  \\
 \hline \hline
    \end{tabular}
\caption{Magnetic moments of the diquark-triquark hidden-bottom pentaquark states in $10_{f}$ representation.  $J_{T}^{P_T} $ corresponds to the angular momentum and parity of triquark, $J_{D}^{P_D}$ is for diquark, $J_{\bar{b}}$ is for anti-bottom, and $J_{l}^{P_{l}}$ is for orbital. The results are listed in unit of nuclear magneton $\mu_N$.}
\label{tab:magmomdtm10f}
\end{table}

\section{Discussion}
\label{sec:discussion}
We systematically study magnetic moments of hidden-bottom pentaquark states with various quantum numbers and spin-couplings. 

In the molecular model, one of the spin configurations of $J^P=\frac{1}{2}^{-}$ is $\frac{1}{2}^{+} \otimes 0^{-} \otimes 0^{+}$. The magnetic moments for this configuration are $\mu=-0.598 ~ \mu_N$ for $8_{1f}$, $\mu=-0.066 ~ \mu_N$ for $8_{2f}$, and $\mu=0.022 ~ \mu_N$ for $10_{f}$ representations. The $\frac{1}{2}^{+} \otimes 0^{-} \otimes 1^{-}$ configuration of $J^P=\frac{1}{2}^{+}$ quantum number has magnetic moment $\mu=0.699 ~ \mu_N$ for $8_{1f}$, $\mu=-0.218 ~ \mu_N$ for $8_{2f}$, and $\mu=-0.522 ~ \mu_N$ for $10_{f}$ representations. A measurement of magnetic moment may present a clue for the representations and therefore quark configurations both in  these quantum numbers. 

In the diquark-diquark-antiquark model, the $S$-wave magnetic moment results have interesting patterns. Some magnetic moments value are same in $8_{1f}$, $8_{2f}$ and $10_f$ representations. For example $1^{+}\otimes1^{+}\otimes{\frac{1}{2}}^{-}\otimes0^{+} $ configuration has same magnetic moment value $\mu=0.066 ~\mu_N$ in $8_{1f}$ and $10_f$ representations. In addition to this, $(1^{+}\otimes 1^{+})_{1} \otimes{\frac{1}{2}}^{-}\otimes 0^{+} $ configuration with $J^P=\frac{3}{2}^{-}$ has magnetic moment $\mu=0.033 ~\mu_N$ in $8_{1f}$ and $10_f$ representations which cannot distinguish inner structure. In the orbitally excited states, both the $\rho$ mode and $\lambda$ mode excitations lead difference in results. 

In the diquark-triquark model, for example $\frac{1}{2}^{-} \otimes 0^{+} \otimes 0^{+}$ of $J^P=\frac{1}{2}^{-}$ has magnetic moment $\mu=-0.642 ~\mu_N$ in $8_{1f}$, $\mu=0.066 ~\mu_N$ in $8_{2f}$ and $\mu=-0.022 ~\mu_N$ in $10_f$ representations, respectively. A possible measurement of magnetic moment may clarify the quark configuration for this state. 

As can be seen in the related tables, the magnetic moment of the hidden-bottom pentaquark states with the same quantum number are different in three models. Even within the same model, $J^P$ quantum numbers may correspond several states with different configurations. As a result, magnetic moments are also different.

It would be interesting to see how three models yield magnetic moments in the same spin quantum number and configuration. For this purpose, we list numerical results of $8_{1f}$ representations of the hidden-bottom pentaquark states in Table \ref{tab:comparison}. As can be seen in table, the results of the molecular model and diquark-triquark model agree at an acceptable level whereas the results of the diquark-diquark-antiquark model are significantly different. The reason for this could be that the configuration of diquark-triquark model is similar to molecular model except the color representation.

\begin{table}[H]
    \centering
    \renewcommand{\arraystretch}{1.25}
    \begin{tabular}{|c|c|c|c|c|c|c|}
        \hline
        $J^P$ &  $^{2s+1}L_j$  & Molecular &  Diquark-diquark-antiquark & Diquark-Triquark   \\\hline\hline
$\frac{1}{2}^-$  \,&\,  $^2S_{\frac{1}{2}}$  \,&\, $-0.598$ \,&\, $-1.928$  \,&\, $-0.642$  \\

  \,&\,   \,&\, $0.864$  \,&\, $0.066$ \,&\, $0.790$ \\

  \,&\,   \,&\, $-0.886$  \,&\, $-0.044$ \,&\, $-0.768$ \\ \hline
  
$\frac{3}{2}^-$   \,&\,  $^4S_{\frac{3}{2}}$  \,&\, $0.399$  \,&\, $-0.864$ \,&\, $0.221$ \\

 \,&\,    \,&\, $-0.997$  \,&\, $0.033$ \,&\, $-0.864$ \\
 
  \,&\,  \,&\, $0.333$  \,&\, $-0.099$ \,&\, $-0.288$ \\ \hline
  
 $\frac{5}{2}^-$   \,&\,  $^6S_{\frac{5}{2}}$  \,&\, $0.237$  \,&\, $2.792$ \,&\, $-1.231$ \\

 \hline \hline
    \end{tabular}
        \caption{Magnetic moments of the $8_{1f}$ hidden-bottom pentaquark states in three different models.  The results are listed in unit of nuclear magneton $\mu_N$.}
          \label{tab:comparison}
\end{table}

\section{Epilogue}
\label{sec:summary}
In recent years, substantial experimental progress has been made in the study of multiquark hadrons. The observation of hidden-charm pentaquark states is a breakthrough in the era of multiquark states. 

In this present work, we systematically study magnetic moments of hidden-bottom pentaquark states in three models: molecular, diquark-diquark-antiquark and diquark-triquark. All three models have different quark and and color configurations. Therefore the interaction in the bound or resonance state is quite different. 

Our results show that magnetic moments of hidden-bottom pentaquark states with different configurations are different. The results clearly show that the same spin configuration has different magnetic moments in same models. The magnetic moment is a guide to explore inner structures of the hadrons. A possible measurement of magnetic moment may display the inner structure of the corresponding state.  Once hidden-bottom pentaquark states are observed experimentally, measurement of magnetic moment can help distinguish their inner structures. We believe that our calculations will be useful for experimental studies.


\bibliography{hb-pentaquark-magnetic}

\begin{thebibliography}{61}%
\makeatletter
\providecommand \@ifxundefined [1]{%
 \@ifx{#1\undefined}
}%
\providecommand \@ifnum [1]{%
 \ifnum #1\expandafter \@firstoftwo
 \else \expandafter \@secondoftwo
 \fi
}%
\providecommand \@ifx [1]{%
 \ifx #1\expandafter \@firstoftwo
 \else \expandafter \@secondoftwo
 \fi
}%
\providecommand \natexlab [1]{#1}%
\providecommand \enquote  [1]{``#1''}%
\providecommand \bibnamefont  [1]{#1}%
\providecommand \bibfnamefont [1]{#1}%
\providecommand \citenamefont [1]{#1}%
\providecommand \href@noop [0]{\@secondoftwo}%
\providecommand \href [0]{\begingroup \@sanitize@url \@href}%
\providecommand \@href[1]{\@@startlink{#1}\@@href}%
\providecommand \@@href[1]{\endgroup#1\@@endlink}%
\providecommand \@sanitize@url [0]{\catcode `\\12\catcode `\$12\catcode
  `\&12\catcode `\#12\catcode `\^12\catcode `\_12\catcode `\%12\relax}%
\providecommand \@@startlink[1]{}%
\providecommand \@@endlink[0]{}%
\providecommand \url  [0]{\begingroup\@sanitize@url \@url }%
\providecommand \@url [1]{\endgroup\@href {#1}{\urlprefix }}%
\providecommand \urlprefix  [0]{URL }%
\providecommand \Eprint [0]{\href }%
\providecommand \doibase [0]{https://doi.org/}%
\providecommand \selectlanguage [0]{\@gobble}%
\providecommand \bibinfo  [0]{\@secondoftwo}%
\providecommand \bibfield  [0]{\@secondoftwo}%
\providecommand \translation [1]{[#1]}%
\providecommand \BibitemOpen [0]{}%
\providecommand \bibitemStop [0]{}%
\providecommand \bibitemNoStop [0]{.\EOS\space}%
\providecommand \EOS [0]{\spacefactor3000\relax}%
\providecommand \BibitemShut  [1]{\csname bibitem#1\endcsname}%
\let\auto@bib@innerbib\@empty
\bibitem [{\citenamefont {Choi}\ \emph {et~al.}(2003)\citenamefont {Choi} \emph
  {et~al.}}]{Belle:2003nnu}%
  \BibitemOpen
  \bibfield  {author} {\bibinfo {author} {\bibfnamefont {S.~K.}\ \bibnamefont
  {Choi}} \emph {et~al.} (\bibinfo {collaboration} {Belle}),\ }\bibfield
  {title} {\bibinfo {title} {{Observation of a narrow charmonium-like state in
  exclusive $B^\pm \to K^\pm \pi^+ \pi^- J/\psi$ decays}},\ }\href
  {https://doi.org/10.1103/PhysRevLett.91.262001} {\bibfield  {journal}
  {\bibinfo  {journal} {Phys. Rev. Lett.}\ }\textbf {\bibinfo {volume} {91}},\
  \bibinfo {pages} {262001} (\bibinfo {year} {2003})},\ \Eprint
  {https://arxiv.org/abs/hep-ex/0309032} {arXiv:hep-ex/0309032} \BibitemShut
  {NoStop}%
\bibitem [{\citenamefont {Aaij}\ \emph {et~al.}(2015)\citenamefont {Aaij} \emph
  {et~al.}}]{LHCb:2015yax}%
  \BibitemOpen
  \bibfield  {author} {\bibinfo {author} {\bibfnamefont {R.}~\bibnamefont
  {Aaij}} \emph {et~al.} (\bibinfo {collaboration} {LHCb}),\ }\bibfield
  {title} {\bibinfo {title} {{Observation of $J/\psi p$ Resonances Consistent
  with Pentaquark States in $\Lambda_b^0 \to J/\psi K^- p$ Decays}},\ }\href
  {https://doi.org/10.1103/PhysRevLett.115.072001} {\bibfield  {journal}
  {\bibinfo  {journal} {Phys. Rev. Lett.}\ }\textbf {\bibinfo {volume} {115}},\
  \bibinfo {pages} {072001} (\bibinfo {year} {2015})},\ \Eprint
  {https://arxiv.org/abs/1507.03414} {arXiv:1507.03414 [hep-ex]} \BibitemShut
  {NoStop}%
\bibitem [{\citenamefont {Aaij}\ \emph {et~al.}(2019)\citenamefont {Aaij} \emph
  {et~al.}}]{LHCb:2019kea}%
  \BibitemOpen
  \bibfield  {author} {\bibinfo {author} {\bibfnamefont {R.}~\bibnamefont
  {Aaij}} \emph {et~al.} (\bibinfo {collaboration} {LHCb}),\ }\bibfield
  {title} {\bibinfo {title} {{Observation of a narrow pentaquark state,
  $P_c(4312)^+$, and of two-peak structure of the $P_c(4450)^+$}},\ }\href
  {https://doi.org/10.1103/PhysRevLett.122.222001} {\bibfield  {journal}
  {\bibinfo  {journal} {Phys. Rev. Lett.}\ }\textbf {\bibinfo {volume} {122}},\
  \bibinfo {pages} {222001} (\bibinfo {year} {2019})},\ \Eprint
  {https://arxiv.org/abs/1904.03947} {arXiv:1904.03947 [hep-ex]} \BibitemShut
  {NoStop}%
\bibitem [{\citenamefont {Aaij}\ \emph {et~al.}(2021)\citenamefont {Aaij} \emph
  {et~al.}}]{LHCb:2020jpq}%
  \BibitemOpen
  \bibfield  {author} {\bibinfo {author} {\bibfnamefont {R.}~\bibnamefont
  {Aaij}} \emph {et~al.} (\bibinfo {collaboration} {LHCb}),\ }\bibfield
  {title} {\bibinfo {title} {{Evidence of a $J/\psi\Lambda$ structure and
  observation of excited $\Xi^-$ states in the $\Xi^-_b \to J/\psi\Lambda K^-$
  decay}},\ }\href {https://doi.org/10.1016/j.scib.2021.02.030} {\bibfield
  {journal} {\bibinfo  {journal} {Sci. Bull.}\ }\textbf {\bibinfo {volume}
  {66}},\ \bibinfo {pages} {1278} (\bibinfo {year} {2021})},\ \Eprint
  {https://arxiv.org/abs/2012.10380} {arXiv:2012.10380 [hep-ex]} \BibitemShut
  {NoStop}%
\bibitem [{\citenamefont {Aaij}\ \emph {et~al.}(2023)\citenamefont {Aaij} \emph
  {et~al.}}]{LHCb:2022ogu}%
  \BibitemOpen
  \bibfield  {author} {\bibinfo {author} {\bibfnamefont {R.}~\bibnamefont
  {Aaij}} \emph {et~al.} (\bibinfo {collaboration} {LHCb}),\ }\bibfield
  {title} {\bibinfo {title} {{Observation of a
  J/\ensuremath{\psi}\ensuremath{\Lambda} Resonance Consistent with a Strange
  Pentaquark Candidate in
  B-\textrightarrow{}J/\ensuremath{\psi}\ensuremath{\Lambda}p\textasciimacron{}
  Decays}},\ }\href {https://doi.org/10.1103/PhysRevLett.131.031901} {\bibfield
   {journal} {\bibinfo  {journal} {Phys. Rev. Lett.}\ }\textbf {\bibinfo
  {volume} {131}},\ \bibinfo {pages} {031901} (\bibinfo {year} {2023})},\
  \Eprint {https://arxiv.org/abs/2210.10346} {arXiv:2210.10346 [hep-ex]}
  \BibitemShut {NoStop}%
\bibitem [{\citenamefont {Lebed}(2015)}]{Lebed:2015tna}%
  \BibitemOpen
  \bibfield  {author} {\bibinfo {author} {\bibfnamefont {R.~F.}\ \bibnamefont
  {Lebed}},\ }\bibfield  {title} {\bibinfo {title} {{The Pentaquark Candidates
  in the Dynamical Diquark Picture}},\ }\href
  {https://doi.org/10.1016/j.physletb.2015.08.032} {\bibfield  {journal}
  {\bibinfo  {journal} {Phys. Lett. B}\ }\textbf {\bibinfo {volume} {749}},\
  \bibinfo {pages} {454} (\bibinfo {year} {2015})},\ \Eprint
  {https://arxiv.org/abs/1507.05867} {arXiv:1507.05867 [hep-ph]} \BibitemShut
  {NoStop}%
\bibitem [{\citenamefont {Li}\ \emph {et~al.}(2015)\citenamefont {Li},
  \citenamefont {He},\ and\ \citenamefont {He}}]{Li:2015gta}%
  \BibitemOpen
  \bibfield  {author} {\bibinfo {author} {\bibfnamefont {G.-N.}\ \bibnamefont
  {Li}}, \bibinfo {author} {\bibfnamefont {X.-G.}\ \bibnamefont {He}},\ and\
  \bibinfo {author} {\bibfnamefont {M.}~\bibnamefont {He}},\ }\bibfield
  {title} {\bibinfo {title} {{Some Predictions of Diquark Model for Hidden
  Charm Pentaquark Discovered at the LHCb}},\ }\href
  {https://doi.org/10.1007/JHEP12(2015)128} {\bibfield  {journal} {\bibinfo
  {journal} {JHEP}\ }\textbf {\bibinfo {volume} {12}},\ \bibinfo {pages}
  {128}},\ \Eprint {https://arxiv.org/abs/1507.08252} {arXiv:1507.08252
  [hep-ph]} \BibitemShut {NoStop}%
\bibitem [{\citenamefont {Maiani}\ \emph {et~al.}(2015)\citenamefont {Maiani},
  \citenamefont {Polosa},\ and\ \citenamefont {Riquer}}]{Maiani:2015vwa}%
  \BibitemOpen
  \bibfield  {author} {\bibinfo {author} {\bibfnamefont {L.}~\bibnamefont
  {Maiani}}, \bibinfo {author} {\bibfnamefont {A.~D.}\ \bibnamefont {Polosa}},\
  and\ \bibinfo {author} {\bibfnamefont {V.}~\bibnamefont {Riquer}},\
  }\bibfield  {title} {\bibinfo {title} {{The New Pentaquarks in the Diquark
  Model}},\ }\href {https://doi.org/10.1016/j.physletb.2015.08.008} {\bibfield
  {journal} {\bibinfo  {journal} {Phys. Lett. B}\ }\textbf {\bibinfo {volume}
  {749}},\ \bibinfo {pages} {289} (\bibinfo {year} {2015})},\ \Eprint
  {https://arxiv.org/abs/1507.04980} {arXiv:1507.04980 [hep-ph]} \BibitemShut
  {NoStop}%
\bibitem [{\citenamefont {Anisovich}\ \emph {et~al.}(2015)\citenamefont
  {Anisovich}, \citenamefont {Matveev}, \citenamefont {Nyiri}, \citenamefont
  {Sarantsev},\ and\ \citenamefont {Semenova}}]{Anisovich:2015cia}%
  \BibitemOpen
  \bibfield  {author} {\bibinfo {author} {\bibfnamefont {V.~V.}\ \bibnamefont
  {Anisovich}}, \bibinfo {author} {\bibfnamefont {M.~A.}\ \bibnamefont
  {Matveev}}, \bibinfo {author} {\bibfnamefont {J.}~\bibnamefont {Nyiri}},
  \bibinfo {author} {\bibfnamefont {A.~V.}\ \bibnamefont {Sarantsev}},\ and\
  \bibinfo {author} {\bibfnamefont {A.~N.}\ \bibnamefont {Semenova}},\
  }\bibfield  {title} {\bibinfo {title} {{Pentaquarks and resonances in the
  $pJ/\psi$ spectrum}},\ }\href@noop {} {\  (\bibinfo {year} {2015})},\ \Eprint
  {https://arxiv.org/abs/1507.07652} {arXiv:1507.07652 [hep-ph]} \BibitemShut
  {NoStop}%
\bibitem [{\citenamefont {Wang}\ and\ \citenamefont
  {Huang}(2016)}]{Wang:2015ava}%
  \BibitemOpen
  \bibfield  {author} {\bibinfo {author} {\bibfnamefont {Z.-G.}\ \bibnamefont
  {Wang}}\ and\ \bibinfo {author} {\bibfnamefont {T.}~\bibnamefont {Huang}},\
  }\bibfield  {title} {\bibinfo {title} {{Analysis of the ${\frac{1}{2}}^{\pm
  }$ pentaquark states in the diquark model with QCD sum rules}},\ }\href
  {https://doi.org/10.1140/epjc/s10052-016-3880-8} {\bibfield  {journal}
  {\bibinfo  {journal} {Eur. Phys. J. C}\ }\textbf {\bibinfo {volume} {76}},\
  \bibinfo {pages} {43} (\bibinfo {year} {2016})},\ \Eprint
  {https://arxiv.org/abs/1508.04189} {arXiv:1508.04189 [hep-ph]} \BibitemShut
  {NoStop}%
\bibitem [{\citenamefont {Wang}(2016{\natexlab{a}})}]{Wang:2015epa}%
  \BibitemOpen
  \bibfield  {author} {\bibinfo {author} {\bibfnamefont {Z.-G.}\ \bibnamefont
  {Wang}},\ }\bibfield  {title} {\bibinfo {title} {{Analysis of $P_c(4380)$ and
  $P_c(4450)$ as pentaquark states in the diquark model with QCD sum rules}},\
  }\href {https://doi.org/10.1140/epjc/s10052-016-3920-4} {\bibfield  {journal}
  {\bibinfo  {journal} {Eur. Phys. J. C}\ }\textbf {\bibinfo {volume} {76}},\
  \bibinfo {pages} {70} (\bibinfo {year} {2016}{\natexlab{a}})},\ \Eprint
  {https://arxiv.org/abs/1508.01468} {arXiv:1508.01468 [hep-ph]} \BibitemShut
  {NoStop}%
\bibitem [{\citenamefont {Wang}(2016{\natexlab{b}})}]{Wang:2015ixb}%
  \BibitemOpen
  \bibfield  {author} {\bibinfo {author} {\bibfnamefont {Z.-G.}\ \bibnamefont
  {Wang}},\ }\bibfield  {title} {\bibinfo {title} {{Analysis of the
  ${\frac{3}{2}}^{\pm}$ pentaquark states in the diquark-diquark-antiquark
  model with QCD sum rules}},\ }\href
  {https://doi.org/10.1016/j.nuclphysb.2016.09.009} {\bibfield  {journal}
  {\bibinfo  {journal} {Nucl. Phys. B}\ }\textbf {\bibinfo {volume} {913}},\
  \bibinfo {pages} {163} (\bibinfo {year} {2016}{\natexlab{b}})},\ \Eprint
  {https://arxiv.org/abs/1512.04763} {arXiv:1512.04763 [hep-ph]} \BibitemShut
  {NoStop}%
\bibitem [{\citenamefont {Ghosh}\ \emph {et~al.}(2017)\citenamefont {Ghosh},
  \citenamefont {Bhattacharya},\ and\ \citenamefont
  {Chakrabarti}}]{Ghosh:2015ksa}%
  \BibitemOpen
  \bibfield  {author} {\bibinfo {author} {\bibfnamefont {R.}~\bibnamefont
  {Ghosh}}, \bibinfo {author} {\bibfnamefont {A.}~\bibnamefont
  {Bhattacharya}},\ and\ \bibinfo {author} {\bibfnamefont {B.}~\bibnamefont
  {Chakrabarti}},\ }\bibfield  {title} {\bibinfo {title} {{A study on
  P$_{c}^{*}$ (4380) and P$_{c}^{*}$ in the quasi particle diquark model}},\
  }\href {https://doi.org/10.1134/S1547477117040100} {\bibfield  {journal}
  {\bibinfo  {journal} {Phys. Part. Nucl. Lett.}\ }\textbf {\bibinfo {volume}
  {14}},\ \bibinfo {pages} {550} (\bibinfo {year} {2017})},\ \Eprint
  {https://arxiv.org/abs/1508.00356} {arXiv:1508.00356 [hep-ph]} \BibitemShut
  {NoStop}%
\bibitem [{\citenamefont {Wang}(2016{\natexlab{c}})}]{Wang:2015wsa}%
  \BibitemOpen
  \bibfield  {author} {\bibinfo {author} {\bibfnamefont {Z.-G.}\ \bibnamefont
  {Wang}},\ }\bibfield  {title} {\bibinfo {title} {{Analysis of the
  ${\frac{1}{2}}^{\pm }$ pentaquark states in the
  diquark\textendash{}diquark\textendash{}antiquark model with QCD sum
  rules}},\ }\href {https://doi.org/10.1140/epjc/s10052-016-3983-2} {\bibfield
  {journal} {\bibinfo  {journal} {Eur. Phys. J. C}\ }\textbf {\bibinfo {volume}
  {76}},\ \bibinfo {pages} {142} (\bibinfo {year} {2016}{\natexlab{c}})},\
  \Eprint {https://arxiv.org/abs/1509.06436} {arXiv:1509.06436 [hep-ph]}
  \BibitemShut {NoStop}%
\bibitem [{\citenamefont {Zhang}\ \emph {et~al.}(2017)\citenamefont {Zhang},
  \citenamefont {Wang},\ and\ \citenamefont {Di}}]{Zhang:2017mmw}%
  \BibitemOpen
  \bibfield  {author} {\bibinfo {author} {\bibfnamefont {J.-X.}\ \bibnamefont
  {Zhang}}, \bibinfo {author} {\bibfnamefont {Z.-G.}\ \bibnamefont {Wang}},\
  and\ \bibinfo {author} {\bibfnamefont {Z.-Y.}\ \bibnamefont {Di}},\
  }\bibfield  {title} {\bibinfo {title} {{Analysis of the $\frac {3}{2}^{\pm }$
  Pentaquark States in the Diquark Model with QCD Sum Rules}},\ }\href
  {https://doi.org/10.5506/APhysPolB.48.2013} {\bibfield  {journal} {\bibinfo
  {journal} {Acta Phys. Polon. B}\ }\textbf {\bibinfo {volume} {48}},\ \bibinfo
  {pages} {2013} (\bibinfo {year} {2017})},\ \Eprint
  {https://arxiv.org/abs/1711.10728} {arXiv:1711.10728 [hep-ph]} \BibitemShut
  {NoStop}%
\bibitem [{\citenamefont {Wang}(2020)}]{Wang:2019got}%
  \BibitemOpen
  \bibfield  {author} {\bibinfo {author} {\bibfnamefont {Z.-G.}\ \bibnamefont
  {Wang}},\ }\bibfield  {title} {\bibinfo {title} {{Analysis of the
  $P_c(4312)$, $P_c(4440)$, $P_c(4457)$ and related hidden-charm pentaquark
  states with QCD sum rules}},\ }\href
  {https://doi.org/10.1142/S0217751X20500037} {\bibfield  {journal} {\bibinfo
  {journal} {Int. J. Mod. Phys. A}\ }\textbf {\bibinfo {volume} {35}},\
  \bibinfo {pages} {2050003} (\bibinfo {year} {2020})},\ \Eprint
  {https://arxiv.org/abs/1905.02892} {arXiv:1905.02892 [hep-ph]} \BibitemShut
  {NoStop}%
\bibitem [{\citenamefont {Wang}\ \emph {et~al.}(2021)\citenamefont {Wang},
  \citenamefont {Wang},\ and\ \citenamefont {Xin}}]{Wang:2020rdh}%
  \BibitemOpen
  \bibfield  {author} {\bibinfo {author} {\bibfnamefont {Z.-G.}\ \bibnamefont
  {Wang}}, \bibinfo {author} {\bibfnamefont {H.-J.}\ \bibnamefont {Wang}},\
  and\ \bibinfo {author} {\bibfnamefont {Q.}~\bibnamefont {Xin}},\ }\bibfield
  {title} {\bibinfo {title} {{The hadronic coupling constants of the lowest
  hidden-charm pentaquark state with the QCD sum rules in rigorous quark-hadron
  duality}},\ }\href {https://doi.org/10.1088/1674-1137/abf13b} {\bibfield
  {journal} {\bibinfo  {journal} {Chin. Phys. C}\ }\textbf {\bibinfo {volume}
  {45}},\ \bibinfo {pages} {063104} (\bibinfo {year} {2021})},\ \Eprint
  {https://arxiv.org/abs/2005.00535} {arXiv:2005.00535 [hep-ph]} \BibitemShut
  {NoStop}%
\bibitem [{\citenamefont {Ali}\ \emph {et~al.}(2021)\citenamefont {Ali},
  \citenamefont {Ahmed}, \citenamefont {Aslam}, \citenamefont {Parkhomenko},\
  and\ \citenamefont {Rehman}}]{Ali:2020vee}%
  \BibitemOpen
  \bibfield  {author} {\bibinfo {author} {\bibfnamefont {A.}~\bibnamefont
  {Ali}}, \bibinfo {author} {\bibfnamefont {I.}~\bibnamefont {Ahmed}}, \bibinfo
  {author} {\bibfnamefont {M.~J.}\ \bibnamefont {Aslam}}, \bibinfo {author}
  {\bibfnamefont {A.}~\bibnamefont {Parkhomenko}},\ and\ \bibinfo {author}
  {\bibfnamefont {A.}~\bibnamefont {Rehman}},\ }\bibfield  {title} {\bibinfo
  {title} {{Interpretation of LHCb Hidden-Charm Pentaquarks within the Compact
  Diquark Model}},\ }\href {https://doi.org/10.22323/1.390.0527} {\bibfield
  {journal} {\bibinfo  {journal} {PoS}\ }\textbf {\bibinfo {volume}
  {ICHEP2020}},\ \bibinfo {pages} {527} (\bibinfo {year} {2021})},\ \Eprint
  {https://arxiv.org/abs/2012.07760} {arXiv:2012.07760 [hep-ph]} \BibitemShut
  {NoStop}%
\bibitem [{\citenamefont {Wang}\ \emph {et~al.}(2016)\citenamefont {Wang},
  \citenamefont {Chen}, \citenamefont {Ma}, \citenamefont {Liu},\ and\
  \citenamefont {Zhu}}]{Wang:2016dzu}%
  \BibitemOpen
  \bibfield  {author} {\bibinfo {author} {\bibfnamefont {G.-J.}\ \bibnamefont
  {Wang}}, \bibinfo {author} {\bibfnamefont {R.}~\bibnamefont {Chen}}, \bibinfo
  {author} {\bibfnamefont {L.}~\bibnamefont {Ma}}, \bibinfo {author}
  {\bibfnamefont {X.}~\bibnamefont {Liu}},\ and\ \bibinfo {author}
  {\bibfnamefont {S.-L.}\ \bibnamefont {Zhu}},\ }\bibfield  {title} {\bibinfo
  {title} {{Magnetic moments of the hidden-charm pentaquark states}},\ }\href
  {https://doi.org/10.1103/PhysRevD.94.094018} {\bibfield  {journal} {\bibinfo
  {journal} {Phys. Rev. D}\ }\textbf {\bibinfo {volume} {94}},\ \bibinfo
  {pages} {094018} (\bibinfo {year} {2016})},\ \Eprint
  {https://arxiv.org/abs/1605.01337} {arXiv:1605.01337 [hep-ph]} \BibitemShut
  {NoStop}%
\bibitem [{\citenamefont {Wang}(2021)}]{Wang:2020eep}%
  \BibitemOpen
  \bibfield  {author} {\bibinfo {author} {\bibfnamefont {Z.-G.}\ \bibnamefont
  {Wang}},\ }\bibfield  {title} {\bibinfo {title} {{Analysis of the
  $P_{cs}(4459)$ as the hidden-charm pentaquark state with QCD sum rules}},\
  }\href {https://doi.org/10.1142/S0217751X21500718} {\bibfield  {journal}
  {\bibinfo  {journal} {Int. J. Mod. Phys. A}\ }\textbf {\bibinfo {volume}
  {36}},\ \bibinfo {pages} {2150071} (\bibinfo {year} {2021})},\ \Eprint
  {https://arxiv.org/abs/2011.05102} {arXiv:2011.05102 [hep-ph]} \BibitemShut
  {NoStop}%
\bibitem [{\citenamefont {Zhu}\ and\ \citenamefont {Qiao}(2016)}]{Zhu:2015bba}%
  \BibitemOpen
  \bibfield  {author} {\bibinfo {author} {\bibfnamefont {R.}~\bibnamefont
  {Zhu}}\ and\ \bibinfo {author} {\bibfnamefont {C.-F.}\ \bibnamefont {Qiao}},\
  }\bibfield  {title} {\bibinfo {title} {{Pentaquark states in a
  diquark\textendash{}triquark model}},\ }\href
  {https://doi.org/10.1016/j.physletb.2016.03.022} {\bibfield  {journal}
  {\bibinfo  {journal} {Phys. Lett. B}\ }\textbf {\bibinfo {volume} {756}},\
  \bibinfo {pages} {259} (\bibinfo {year} {2016})},\ \Eprint
  {https://arxiv.org/abs/1510.08693} {arXiv:1510.08693 [hep-ph]} \BibitemShut
  {NoStop}%
\bibitem [{\citenamefont {Chen}\ \emph
  {et~al.}(2015{\natexlab{a}})\citenamefont {Chen}, \citenamefont {Liu},
  \citenamefont {Li},\ and\ \citenamefont {Zhu}}]{Chen:2015loa}%
  \BibitemOpen
  \bibfield  {author} {\bibinfo {author} {\bibfnamefont {R.}~\bibnamefont
  {Chen}}, \bibinfo {author} {\bibfnamefont {X.}~\bibnamefont {Liu}}, \bibinfo
  {author} {\bibfnamefont {X.-Q.}\ \bibnamefont {Li}},\ and\ \bibinfo {author}
  {\bibfnamefont {S.-L.}\ \bibnamefont {Zhu}},\ }\bibfield  {title} {\bibinfo
  {title} {{Identifying exotic hidden-charm pentaquarks}},\ }\href
  {https://doi.org/10.1103/PhysRevLett.115.132002} {\bibfield  {journal}
  {\bibinfo  {journal} {Phys. Rev. Lett.}\ }\textbf {\bibinfo {volume} {115}},\
  \bibinfo {pages} {132002} (\bibinfo {year} {2015}{\natexlab{a}})},\ \Eprint
  {https://arxiv.org/abs/1507.03704} {arXiv:1507.03704 [hep-ph]} \BibitemShut
  {NoStop}%
\bibitem [{\citenamefont {Chen}\ \emph
  {et~al.}(2015{\natexlab{b}})\citenamefont {Chen}, \citenamefont {Chen},
  \citenamefont {Liu}, \citenamefont {Steele},\ and\ \citenamefont
  {Zhu}}]{Chen:2015moa}%
  \BibitemOpen
  \bibfield  {author} {\bibinfo {author} {\bibfnamefont {H.-X.}\ \bibnamefont
  {Chen}}, \bibinfo {author} {\bibfnamefont {W.}~\bibnamefont {Chen}}, \bibinfo
  {author} {\bibfnamefont {X.}~\bibnamefont {Liu}}, \bibinfo {author}
  {\bibfnamefont {T.~G.}\ \bibnamefont {Steele}},\ and\ \bibinfo {author}
  {\bibfnamefont {S.-L.}\ \bibnamefont {Zhu}},\ }\bibfield  {title} {\bibinfo
  {title} {{Towards exotic hidden-charm pentaquarks in QCD}},\ }\href
  {https://doi.org/10.1103/PhysRevLett.115.172001} {\bibfield  {journal}
  {\bibinfo  {journal} {Phys. Rev. Lett.}\ }\textbf {\bibinfo {volume} {115}},\
  \bibinfo {pages} {172001} (\bibinfo {year} {2015}{\natexlab{b}})},\ \Eprint
  {https://arxiv.org/abs/1507.03717} {arXiv:1507.03717 [hep-ph]} \BibitemShut
  {NoStop}%
\bibitem [{\citenamefont {He}(2016)}]{He:2015cea}%
  \BibitemOpen
  \bibfield  {author} {\bibinfo {author} {\bibfnamefont {J.}~\bibnamefont
  {He}},\ }\bibfield  {title} {\bibinfo {title} {{$\bar{D}\Sigma^*_c$ and
  $\bar{D}^*\Sigma_c$ interactions and the LHCb hidden-charmed pentaquarks}},\
  }\href {https://doi.org/10.1016/j.physletb.2015.12.071} {\bibfield  {journal}
  {\bibinfo  {journal} {Phys. Lett. B}\ }\textbf {\bibinfo {volume} {753}},\
  \bibinfo {pages} {547} (\bibinfo {year} {2016})},\ \Eprint
  {https://arxiv.org/abs/1507.05200} {arXiv:1507.05200 [hep-ph]} \BibitemShut
  {NoStop}%
\bibitem [{\citenamefont {Mei\ss{}ner}\ and\ \citenamefont
  {Oller}(2015)}]{Meissner:2015mza}%
  \BibitemOpen
  \bibfield  {author} {\bibinfo {author} {\bibfnamefont {U.-G.}\ \bibnamefont
  {Mei\ss{}ner}}\ and\ \bibinfo {author} {\bibfnamefont {J.~A.}\ \bibnamefont
  {Oller}},\ }\bibfield  {title} {\bibinfo {title} {{Testing the $\chi_{c1}\,
  p$ composite nature of the $P_c(4450)$}},\ }\href
  {https://doi.org/10.1016/j.physletb.2015.10.015} {\bibfield  {journal}
  {\bibinfo  {journal} {Phys. Lett. B}\ }\textbf {\bibinfo {volume} {751}},\
  \bibinfo {pages} {59} (\bibinfo {year} {2015})},\ \Eprint
  {https://arxiv.org/abs/1507.07478} {arXiv:1507.07478 [hep-ph]} \BibitemShut
  {NoStop}%
\bibitem [{\citenamefont {Roca}\ \emph {et~al.}(2015)\citenamefont {Roca},
  \citenamefont {Nieves},\ and\ \citenamefont {Oset}}]{Roca:2015dva}%
  \BibitemOpen
  \bibfield  {author} {\bibinfo {author} {\bibfnamefont {L.}~\bibnamefont
  {Roca}}, \bibinfo {author} {\bibfnamefont {J.}~\bibnamefont {Nieves}},\ and\
  \bibinfo {author} {\bibfnamefont {E.}~\bibnamefont {Oset}},\ }\bibfield
  {title} {\bibinfo {title} {{LHCb pentaquark as a
  $\bar{D}^*\Sigma_c-\bar{D}^*\Sigma_c^*$ molecular state}},\ }\href
  {https://doi.org/10.1103/PhysRevD.92.094003} {\bibfield  {journal} {\bibinfo
  {journal} {Phys. Rev. D}\ }\textbf {\bibinfo {volume} {92}},\ \bibinfo
  {pages} {094003} (\bibinfo {year} {2015})},\ \Eprint
  {https://arxiv.org/abs/1507.04249} {arXiv:1507.04249 [hep-ph]} \BibitemShut
  {NoStop}%
\bibitem [{\citenamefont {Wang}(2019)}]{Wang:2018waa}%
  \BibitemOpen
  \bibfield  {author} {\bibinfo {author} {\bibfnamefont {Z.-G.}\ \bibnamefont
  {Wang}},\ }\bibfield  {title} {\bibinfo {title} {{Analysis of the
  $\bar{D}\Sigma_c$, $\bar{D}\Sigma_c^*$, $\bar{D}^{*}\Sigma_c$ and
  $\bar{D}^{*}\Sigma_c^*$ pentaquark molecular states with QCD sum rules}},\
  }\href {https://doi.org/10.1142/S0217751X19500970} {\bibfield  {journal}
  {\bibinfo  {journal} {Int. J. Mod. Phys. A}\ }\textbf {\bibinfo {volume}
  {34}},\ \bibinfo {pages} {1950097} (\bibinfo {year} {2019})},\ \Eprint
  {https://arxiv.org/abs/1806.10384} {arXiv:1806.10384 [hep-ph]} \BibitemShut
  {NoStop}%
\bibitem [{\citenamefont {Wang}\ and\ \citenamefont
  {Wang}(2020)}]{Wang:2019hyc}%
  \BibitemOpen
  \bibfield  {author} {\bibinfo {author} {\bibfnamefont {Z.-G.}\ \bibnamefont
  {Wang}}\ and\ \bibinfo {author} {\bibfnamefont {X.}~\bibnamefont {Wang}},\
  }\bibfield  {title} {\bibinfo {title} {{Analysis of the strong decays of the
  $P_c(4312)$ as a pentaquark molecular state with QCD sum rules}},\ }\href
  {https://doi.org/10.1088/1674-1137/ababf7} {\bibfield  {journal} {\bibinfo
  {journal} {Chin. Phys. C}\ }\textbf {\bibinfo {volume} {44}},\ \bibinfo
  {pages} {103102} (\bibinfo {year} {2020})},\ \Eprint
  {https://arxiv.org/abs/1907.04582} {arXiv:1907.04582 [hep-ph]} \BibitemShut
  {NoStop}%
\bibitem [{\citenamefont {Chen}\ \emph {et~al.}(2021)\citenamefont {Chen},
  \citenamefont {Chen}, \citenamefont {Liu},\ and\ \citenamefont
  {Liu}}]{Chen:2020uif}%
  \BibitemOpen
  \bibfield  {author} {\bibinfo {author} {\bibfnamefont {H.-X.}\ \bibnamefont
  {Chen}}, \bibinfo {author} {\bibfnamefont {W.}~\bibnamefont {Chen}}, \bibinfo
  {author} {\bibfnamefont {X.}~\bibnamefont {Liu}},\ and\ \bibinfo {author}
  {\bibfnamefont {X.-H.}\ \bibnamefont {Liu}},\ }\bibfield  {title} {\bibinfo
  {title} {{Establishing the first hidden-charm pentaquark with strangeness}},\
  }\href {https://doi.org/10.1140/epjc/s10052-021-09196-4} {\bibfield
  {journal} {\bibinfo  {journal} {Eur. Phys. J. C}\ }\textbf {\bibinfo {volume}
  {81}},\ \bibinfo {pages} {409} (\bibinfo {year} {2021})},\ \Eprint
  {https://arxiv.org/abs/2011.01079} {arXiv:2011.01079 [hep-ph]} \BibitemShut
  {NoStop}%
\bibitem [{\citenamefont {Peng}\ \emph {et~al.}(2021)\citenamefont {Peng},
  \citenamefont {Yan}, \citenamefont {S\'anchez~S\'anchez},\ and\ \citenamefont
  {Valderrama}}]{Peng:2020hql}%
  \BibitemOpen
  \bibfield  {author} {\bibinfo {author} {\bibfnamefont {F.-Z.}\ \bibnamefont
  {Peng}}, \bibinfo {author} {\bibfnamefont {M.-J.}\ \bibnamefont {Yan}},
  \bibinfo {author} {\bibfnamefont {M.}~\bibnamefont {S\'anchez~S\'anchez}},\
  and\ \bibinfo {author} {\bibfnamefont {M.~P.}\ \bibnamefont {Valderrama}},\
  }\bibfield  {title} {\bibinfo {title} {{The $P_{cs}(4459)$ pentaquark from a
  combined effective field theory and phenomenological perspective}},\ }\href
  {https://doi.org/10.1140/epjc/s10052-021-09416-x} {\bibfield  {journal}
  {\bibinfo  {journal} {Eur. Phys. J. C}\ }\textbf {\bibinfo {volume} {81}},\
  \bibinfo {pages} {666} (\bibinfo {year} {2021})},\ \Eprint
  {https://arxiv.org/abs/2011.01915} {arXiv:2011.01915 [hep-ph]} \BibitemShut
  {NoStop}%
\bibitem [{\citenamefont {Chen}(2021)}]{Chen:2020kco}%
  \BibitemOpen
  \bibfield  {author} {\bibinfo {author} {\bibfnamefont {R.}~\bibnamefont
  {Chen}},\ }\bibfield  {title} {\bibinfo {title} {{Can the newly reported
  $P_{cs}(4459)$ be a strange hidden-charm $\Xi_c\bar D^*$ molecular
  pentaquark?}},\ }\href {https://doi.org/10.1103/PhysRevD.103.054007}
  {\bibfield  {journal} {\bibinfo  {journal} {Phys. Rev. D}\ }\textbf {\bibinfo
  {volume} {103}},\ \bibinfo {pages} {054007} (\bibinfo {year} {2021})},\
  \Eprint {https://arxiv.org/abs/2011.07214} {arXiv:2011.07214 [hep-ph]}
  \BibitemShut {NoStop}%
\bibitem [{\citenamefont {Chen}(2022)}]{Chen:2020opr}%
  \BibitemOpen
  \bibfield  {author} {\bibinfo {author} {\bibfnamefont {H.-X.}\ \bibnamefont
  {Chen}},\ }\bibfield  {title} {\bibinfo {title} {{Hidden-charm pentaquark
  states through current algebra: from their production to decay *}},\ }\href
  {https://doi.org/10.1088/1674-1137/ac6ed2} {\bibfield  {journal} {\bibinfo
  {journal} {Chin. Phys. C}\ }\textbf {\bibinfo {volume} {46}},\ \bibinfo
  {pages} {093105} (\bibinfo {year} {2022})},\ \Eprint
  {https://arxiv.org/abs/2011.07187} {arXiv:2011.07187 [hep-ph]} \BibitemShut
  {NoStop}%
\bibitem [{\citenamefont {Wang}\ and\ \citenamefont
  {Xin}(2021)}]{Wang:2021itn}%
  \BibitemOpen
  \bibfield  {author} {\bibinfo {author} {\bibfnamefont {Z.-G.}\ \bibnamefont
  {Wang}}\ and\ \bibinfo {author} {\bibfnamefont {Q.}~\bibnamefont {Xin}},\
  }\bibfield  {title} {\bibinfo {title} {{Analysis of hidden-charm pentaquark
  molecular states with and without strangeness via the QCD sum rules *}},\
  }\href {https://doi.org/10.1088/1674-1137/ac2a1d} {\bibfield  {journal}
  {\bibinfo  {journal} {Chin. Phys. C}\ }\textbf {\bibinfo {volume} {45}},\
  \bibinfo {pages} {123105} (\bibinfo {year} {2021})},\ \Eprint
  {https://arxiv.org/abs/2103.08239} {arXiv:2103.08239 [hep-ph]} \BibitemShut
  {NoStop}%
\bibitem [{\citenamefont {Wang}\ \emph {et~al.}(2023)\citenamefont {Wang},
  \citenamefont {Chen}, \citenamefont {Meng},\ and\ \citenamefont
  {Zhu}}]{Wang:2023eng}%
  \BibitemOpen
  \bibfield  {author} {\bibinfo {author} {\bibfnamefont {B.}~\bibnamefont
  {Wang}}, \bibinfo {author} {\bibfnamefont {K.}~\bibnamefont {Chen}}, \bibinfo
  {author} {\bibfnamefont {L.}~\bibnamefont {Meng}},\ and\ \bibinfo {author}
  {\bibfnamefont {S.-L.}\ \bibnamefont {Zhu}},\ }\bibfield  {title} {\bibinfo
  {title} {{Spectrum of the molecular pentaquarks}},\ }\href@noop {} {\
  (\bibinfo {year} {2023})},\ \Eprint {https://arxiv.org/abs/2312.13591}
  {arXiv:2312.13591 [hep-ph]} \BibitemShut {NoStop}%
\bibitem [{\citenamefont {Li}\ \emph {et~al.}(2024)\citenamefont {Li},
  \citenamefont {Guo}, \citenamefont {Lei},\ and\ \citenamefont
  {Gao}}]{Li:2024wxr}%
  \BibitemOpen
  \bibfield  {author} {\bibinfo {author} {\bibfnamefont {H.-S.}\ \bibnamefont
  {Li}}, \bibinfo {author} {\bibfnamefont {F.}~\bibnamefont {Guo}}, \bibinfo
  {author} {\bibfnamefont {Y.-D.}\ \bibnamefont {Lei}},\ and\ \bibinfo {author}
  {\bibfnamefont {F.}~\bibnamefont {Gao}},\ }\bibfield  {title} {\bibinfo
  {title} {{Magnetic moments and axial charges of the octet hidden-charm
  molecular pentaquark family}},\ }\href@noop {} {\  (\bibinfo {year}
  {2024})},\ \Eprint {https://arxiv.org/abs/2401.14767} {arXiv:2401.14767
  [hep-ph]} \BibitemShut {NoStop}%
\bibitem [{\citenamefont {Li}(2024)}]{Li:2024jlq}%
  \BibitemOpen
  \bibfield  {author} {\bibinfo {author} {\bibfnamefont {H.-S.}\ \bibnamefont
  {Li}},\ }\bibfield  {title} {\bibinfo {title} {{Ab initio calculation of
  molecular pentaquark magnetic moments in heavy pentaquark chiral perturbation
  theory}},\ }\href@noop {} {\  (\bibinfo {year} {2024})},\ \Eprint
  {https://arxiv.org/abs/2401.14759} {arXiv:2401.14759 [hep-ph]} \BibitemShut
  {NoStop}%
\bibitem [{\citenamefont {Scoccola}\ \emph {et~al.}(2015)\citenamefont
  {Scoccola}, \citenamefont {Riska},\ and\ \citenamefont
  {Rho}}]{Scoccola:2015nia}%
  \BibitemOpen
  \bibfield  {author} {\bibinfo {author} {\bibfnamefont {N.~N.}\ \bibnamefont
  {Scoccola}}, \bibinfo {author} {\bibfnamefont {D.~O.}\ \bibnamefont
  {Riska}},\ and\ \bibinfo {author} {\bibfnamefont {M.}~\bibnamefont {Rho}},\
  }\bibfield  {title} {\bibinfo {title} {{Pentaquark candidates P$_c^+$(4380)
  and P$_c^+$(4450) within the soliton picture of baryons}},\ }\href
  {https://doi.org/10.1103/PhysRevD.92.051501} {\bibfield  {journal} {\bibinfo
  {journal} {Phys. Rev. D}\ }\textbf {\bibinfo {volume} {92}},\ \bibinfo
  {pages} {051501} (\bibinfo {year} {2015})},\ \Eprint
  {https://arxiv.org/abs/1508.01172} {arXiv:1508.01172 [hep-ph]} \BibitemShut
  {NoStop}%
\bibitem [{\citenamefont {Carlson}\ \emph {et~al.}(1988)\citenamefont
  {Carlson}, \citenamefont {Heller},\ and\ \citenamefont
  {Tjon}}]{Carlson:1987hh}%
  \BibitemOpen
  \bibfield  {author} {\bibinfo {author} {\bibfnamefont {J.}~\bibnamefont
  {Carlson}}, \bibinfo {author} {\bibfnamefont {L.}~\bibnamefont {Heller}},\
  and\ \bibinfo {author} {\bibfnamefont {J.~A.}\ \bibnamefont {Tjon}},\
  }\bibfield  {title} {\bibinfo {title} {{Stability of Dimesons}},\ }\href
  {https://doi.org/10.1103/PhysRevD.37.744} {\bibfield  {journal} {\bibinfo
  {journal} {Phys. Rev. D}\ }\textbf {\bibinfo {volume} {37}},\ \bibinfo
  {pages} {744} (\bibinfo {year} {1988})}\BibitemShut {NoStop}%
\bibitem [{\citenamefont {Manohar}\ and\ \citenamefont
  {Wise}(1993)}]{Manohar:1992nd}%
  \BibitemOpen
  \bibfield  {author} {\bibinfo {author} {\bibfnamefont {A.~V.}\ \bibnamefont
  {Manohar}}\ and\ \bibinfo {author} {\bibfnamefont {M.~B.}\ \bibnamefont
  {Wise}},\ }\bibfield  {title} {\bibinfo {title} {{Exotic Q Q anti-q anti-q
  states in QCD}},\ }\href {https://doi.org/10.1016/0550-3213(93)90614-U}
  {\bibfield  {journal} {\bibinfo  {journal} {Nucl. Phys. B}\ }\textbf
  {\bibinfo {volume} {399}},\ \bibinfo {pages} {17} (\bibinfo {year} {1993})},\
  \Eprint {https://arxiv.org/abs/hep-ph/9212236} {arXiv:hep-ph/9212236}
  \BibitemShut {NoStop}%
\bibitem [{\citenamefont {Aaij}\ \emph
  {et~al.}(2022{\natexlab{a}})\citenamefont {Aaij} \emph
  {et~al.}}]{LHCb:2021vvq}%
  \BibitemOpen
  \bibfield  {author} {\bibinfo {author} {\bibfnamefont {R.}~\bibnamefont
  {Aaij}} \emph {et~al.} (\bibinfo {collaboration} {LHCb}),\ }\bibfield
  {title} {\bibinfo {title} {{Observation of an exotic narrow doubly charmed
  tetraquark}},\ }\href {https://doi.org/10.1038/s41567-022-01614-y} {\bibfield
   {journal} {\bibinfo  {journal} {Nature Phys.}\ }\textbf {\bibinfo {volume}
  {18}},\ \bibinfo {pages} {751} (\bibinfo {year} {2022}{\natexlab{a}})},\
  \Eprint {https://arxiv.org/abs/2109.01038} {arXiv:2109.01038 [hep-ex]}
  \BibitemShut {NoStop}%
\bibitem [{\citenamefont {Aaij}\ \emph
  {et~al.}(2022{\natexlab{b}})\citenamefont {Aaij} \emph
  {et~al.}}]{LHCb:2021auc}%
  \BibitemOpen
  \bibfield  {author} {\bibinfo {author} {\bibfnamefont {R.}~\bibnamefont
  {Aaij}} \emph {et~al.} (\bibinfo {collaboration} {LHCb}),\ }\bibfield
  {title} {\bibinfo {title} {{Study of the doubly charmed tetraquark
  $T_{cc}^{+}$}},\ }\href {https://doi.org/10.1038/s41467-022-30206-w}
  {\bibfield  {journal} {\bibinfo  {journal} {Nature Commun.}\ }\textbf
  {\bibinfo {volume} {13}},\ \bibinfo {pages} {3351} (\bibinfo {year}
  {2022}{\natexlab{b}})},\ \Eprint {https://arxiv.org/abs/2109.01056}
  {arXiv:2109.01056 [hep-ex]} \BibitemShut {NoStop}%
\bibitem [{\citenamefont {Francis}\ \emph {et~al.}(2017)\citenamefont
  {Francis}, \citenamefont {Hudspith}, \citenamefont {Lewis},\ and\
  \citenamefont {Maltman}}]{Francis:2016hui}%
  \BibitemOpen
  \bibfield  {author} {\bibinfo {author} {\bibfnamefont {A.}~\bibnamefont
  {Francis}}, \bibinfo {author} {\bibfnamefont {R.~J.}\ \bibnamefont
  {Hudspith}}, \bibinfo {author} {\bibfnamefont {R.}~\bibnamefont {Lewis}},\
  and\ \bibinfo {author} {\bibfnamefont {K.}~\bibnamefont {Maltman}},\
  }\bibfield  {title} {\bibinfo {title} {{Lattice Prediction for Deeply Bound
  Doubly Heavy Tetraquarks}},\ }\href
  {https://doi.org/10.1103/PhysRevLett.118.142001} {\bibfield  {journal}
  {\bibinfo  {journal} {Phys. Rev. Lett.}\ }\textbf {\bibinfo {volume} {118}},\
  \bibinfo {pages} {142001} (\bibinfo {year} {2017})},\ \Eprint
  {https://arxiv.org/abs/1607.05214} {arXiv:1607.05214 [hep-lat]} \BibitemShut
  {NoStop}%
\bibitem [{\citenamefont {Junnarkar}\ \emph {et~al.}(2019)\citenamefont
  {Junnarkar}, \citenamefont {Mathur},\ and\ \citenamefont
  {Padmanath}}]{Junnarkar:2018twb}%
  \BibitemOpen
  \bibfield  {author} {\bibinfo {author} {\bibfnamefont {P.}~\bibnamefont
  {Junnarkar}}, \bibinfo {author} {\bibfnamefont {N.}~\bibnamefont {Mathur}},\
  and\ \bibinfo {author} {\bibfnamefont {M.}~\bibnamefont {Padmanath}},\
  }\bibfield  {title} {\bibinfo {title} {{Study of doubly heavy tetraquarks in
  Lattice QCD}},\ }\href {https://doi.org/10.1103/PhysRevD.99.034507}
  {\bibfield  {journal} {\bibinfo  {journal} {Phys. Rev. D}\ }\textbf {\bibinfo
  {volume} {99}},\ \bibinfo {pages} {034507} (\bibinfo {year} {2019})},\
  \Eprint {https://arxiv.org/abs/1810.12285} {arXiv:1810.12285 [hep-lat]}
  \BibitemShut {NoStop}%
\bibitem [{\citenamefont {Leskovec}\ \emph {et~al.}(2019)\citenamefont
  {Leskovec}, \citenamefont {Meinel}, \citenamefont {Pflaumer},\ and\
  \citenamefont {Wagner}}]{Leskovec:2019ioa}%
  \BibitemOpen
  \bibfield  {author} {\bibinfo {author} {\bibfnamefont {L.}~\bibnamefont
  {Leskovec}}, \bibinfo {author} {\bibfnamefont {S.}~\bibnamefont {Meinel}},
  \bibinfo {author} {\bibfnamefont {M.}~\bibnamefont {Pflaumer}},\ and\
  \bibinfo {author} {\bibfnamefont {M.}~\bibnamefont {Wagner}},\ }\bibfield
  {title} {\bibinfo {title} {{Lattice QCD investigation of a doubly-bottom
  $\bar{b} \bar{b} u d$ tetraquark with quantum numbers $I(J^P) = 0(1^+)$}},\
  }\href {https://doi.org/10.1103/PhysRevD.100.014503} {\bibfield  {journal}
  {\bibinfo  {journal} {Phys. Rev. D}\ }\textbf {\bibinfo {volume} {100}},\
  \bibinfo {pages} {014503} (\bibinfo {year} {2019})},\ \Eprint
  {https://arxiv.org/abs/1904.04197} {arXiv:1904.04197 [hep-lat]} \BibitemShut
  {NoStop}%
\bibitem [{\citenamefont {Mohanta}\ and\ \citenamefont
  {Basak}(2020)}]{Mohanta:2020eed}%
  \BibitemOpen
  \bibfield  {author} {\bibinfo {author} {\bibfnamefont {P.}~\bibnamefont
  {Mohanta}}\ and\ \bibinfo {author} {\bibfnamefont {S.}~\bibnamefont
  {Basak}},\ }\bibfield  {title} {\bibinfo {title} {{Construction of
  $bb\bar{u}\bar{d}$ tetraquark states on lattice with NRQCD bottom and HISQ up
  and down quarks}},\ }\href {https://doi.org/10.1103/PhysRevD.102.094516}
  {\bibfield  {journal} {\bibinfo  {journal} {Phys. Rev. D}\ }\textbf {\bibinfo
  {volume} {102}},\ \bibinfo {pages} {094516} (\bibinfo {year} {2020})},\
  \Eprint {https://arxiv.org/abs/2008.11146} {arXiv:2008.11146 [hep-lat]}
  \BibitemShut {NoStop}%
\bibitem [{\citenamefont {Alexandrou}\ \emph {et~al.}(2023)\citenamefont
  {Alexandrou}, \citenamefont {Finkenrath}, \citenamefont {Leontiou},
  \citenamefont {Meinel}, \citenamefont {Pflaumer},\ and\ \citenamefont
  {Wagner}}]{Alexandrou:2023cqg}%
  \BibitemOpen
  \bibfield  {author} {\bibinfo {author} {\bibfnamefont {C.}~\bibnamefont
  {Alexandrou}}, \bibinfo {author} {\bibfnamefont {J.}~\bibnamefont
  {Finkenrath}}, \bibinfo {author} {\bibfnamefont {T.}~\bibnamefont
  {Leontiou}}, \bibinfo {author} {\bibfnamefont {S.}~\bibnamefont {Meinel}},
  \bibinfo {author} {\bibfnamefont {M.}~\bibnamefont {Pflaumer}},\ and\
  \bibinfo {author} {\bibfnamefont {M.}~\bibnamefont {Wagner}},\ }\bibfield
  {title} {\bibinfo {title} {{Evidence for shallow bound states and hints for
  broad resonances with quark content $\bar{b}\bar{c}ud$ in $B$-$\bar{D}$ and
  $B^*$-$\bar{D}$ scattering from lattice QCD}},\ }\href@noop {} {\  (\bibinfo
  {year} {2023})},\ \Eprint {https://arxiv.org/abs/2312.02925}
  {arXiv:2312.02925 [hep-lat]} \BibitemShut {NoStop}%
\bibitem [{\citenamefont {Shimizu}\ \emph {et~al.}(2016)\citenamefont
  {Shimizu}, \citenamefont {Suenaga},\ and\ \citenamefont
  {Harada}}]{Shimizu:2016rrd}%
  \BibitemOpen
  \bibfield  {author} {\bibinfo {author} {\bibfnamefont {Y.}~\bibnamefont
  {Shimizu}}, \bibinfo {author} {\bibfnamefont {D.}~\bibnamefont {Suenaga}},\
  and\ \bibinfo {author} {\bibfnamefont {M.}~\bibnamefont {Harada}},\
  }\bibfield  {title} {\bibinfo {title} {{Coupled channel analysis of molecule
  picture of $P_{c}(4380)$}},\ }\href
  {https://doi.org/10.1103/PhysRevD.93.114003} {\bibfield  {journal} {\bibinfo
  {journal} {Phys. Rev. D}\ }\textbf {\bibinfo {volume} {93}},\ \bibinfo
  {pages} {114003} (\bibinfo {year} {2016})},\ \Eprint
  {https://arxiv.org/abs/1603.02376} {arXiv:1603.02376 [hep-ph]} \BibitemShut
  {NoStop}%
\bibitem [{\citenamefont {Liu}\ and\ \citenamefont
  {Zahed}(2017)}]{Liu:2017xzo}%
  \BibitemOpen
  \bibfield  {author} {\bibinfo {author} {\bibfnamefont {Y.}~\bibnamefont
  {Liu}}\ and\ \bibinfo {author} {\bibfnamefont {I.}~\bibnamefont {Zahed}},\
  }\bibfield  {title} {\bibinfo {title} {{Heavy Baryons and their Exotics from
  Instantons in Holographic QCD}},\ }\href
  {https://doi.org/10.1103/PhysRevD.95.116012} {\bibfield  {journal} {\bibinfo
  {journal} {Phys. Rev. D}\ }\textbf {\bibinfo {volume} {95}},\ \bibinfo
  {pages} {116012} (\bibinfo {year} {2017})},\ \Eprint
  {https://arxiv.org/abs/1704.03412} {arXiv:1704.03412 [hep-ph]} \BibitemShut
  {NoStop}%
\bibitem [{\citenamefont {Sakai}\ and\ \citenamefont
  {Sugimoto}(2005)}]{Sakai:2004cn}%
  \BibitemOpen
  \bibfield  {author} {\bibinfo {author} {\bibfnamefont {T.}~\bibnamefont
  {Sakai}}\ and\ \bibinfo {author} {\bibfnamefont {S.}~\bibnamefont
  {Sugimoto}},\ }\bibfield  {title} {\bibinfo {title} {{Low energy hadron
  physics in holographic QCD}},\ }\href {https://doi.org/10.1143/PTP.113.843}
  {\bibfield  {journal} {\bibinfo  {journal} {Prog. Theor. Phys.}\ }\textbf
  {\bibinfo {volume} {113}},\ \bibinfo {pages} {843} (\bibinfo {year}
  {2005})},\ \Eprint {https://arxiv.org/abs/hep-th/0412141}
  {arXiv:hep-th/0412141} \BibitemShut {NoStop}%
\bibitem [{\citenamefont {Azizi}\ \emph {et~al.}(2017)\citenamefont {Azizi},
  \citenamefont {Sarac},\ and\ \citenamefont {Sundu}}]{Azizi:2017bgs}%
  \BibitemOpen
  \bibfield  {author} {\bibinfo {author} {\bibfnamefont {K.}~\bibnamefont
  {Azizi}}, \bibinfo {author} {\bibfnamefont {Y.}~\bibnamefont {Sarac}},\ and\
  \bibinfo {author} {\bibfnamefont {H.}~\bibnamefont {Sundu}},\ }\bibfield
  {title} {\bibinfo {title} {{Hidden Bottom Pentaquark States with Spin 3/2 and
  5/2}},\ }\href {https://doi.org/10.1103/PhysRevD.96.094030} {\bibfield
  {journal} {\bibinfo  {journal} {Phys. Rev. D}\ }\textbf {\bibinfo {volume}
  {96}},\ \bibinfo {pages} {094030} (\bibinfo {year} {2017})},\ \Eprint
  {https://arxiv.org/abs/1707.01248} {arXiv:1707.01248 [hep-ph]} \BibitemShut
  {NoStop}%
\bibitem [{\citenamefont {Yang}\ \emph {et~al.}(2019)\citenamefont {Yang},
  \citenamefont {Ping},\ and\ \citenamefont {Segovia}}]{Yang:2018oqd}%
  \BibitemOpen
  \bibfield  {author} {\bibinfo {author} {\bibfnamefont {G.}~\bibnamefont
  {Yang}}, \bibinfo {author} {\bibfnamefont {J.}~\bibnamefont {Ping}},\ and\
  \bibinfo {author} {\bibfnamefont {J.}~\bibnamefont {Segovia}},\ }\bibfield
  {title} {\bibinfo {title} {{Hidden-bottom pentaquarks}},\ }\href
  {https://doi.org/10.1103/PhysRevD.99.014035} {\bibfield  {journal} {\bibinfo
  {journal} {Phys. Rev. D}\ }\textbf {\bibinfo {volume} {99}},\ \bibinfo
  {pages} {014035} (\bibinfo {year} {2019})},\ \Eprint
  {https://arxiv.org/abs/1809.06193} {arXiv:1809.06193 [hep-ph]} \BibitemShut
  {NoStop}%
\bibitem [{\citenamefont {Zhu}\ \emph {et~al.}(2020)\citenamefont {Zhu},
  \citenamefont {Kong}, \citenamefont {Liu},\ and\ \citenamefont
  {He}}]{Zhu:2020vto}%
  \BibitemOpen
  \bibfield  {author} {\bibinfo {author} {\bibfnamefont {J.-T.}\ \bibnamefont
  {Zhu}}, \bibinfo {author} {\bibfnamefont {S.-Y.}\ \bibnamefont {Kong}},
  \bibinfo {author} {\bibfnamefont {Y.}~\bibnamefont {Liu}},\ and\ \bibinfo
  {author} {\bibfnamefont {J.}~\bibnamefont {He}},\ }\bibfield  {title}
  {\bibinfo {title} {{Hidden-bottom molecular states from $\Sigma
  ^{(*)}_bB^{(*)}-\Lambda _bB^{(*)}$ interaction}},\ }\href
  {https://doi.org/10.1140/epjc/s10052-020-8410-z} {\bibfield  {journal}
  {\bibinfo  {journal} {Eur. Phys. J. C}\ }\textbf {\bibinfo {volume} {80}},\
  \bibinfo {pages} {1016} (\bibinfo {year} {2020})},\ \Eprint
  {https://arxiv.org/abs/2007.07596} {arXiv:2007.07596 [hep-ph]} \BibitemShut
  {NoStop}%
\bibitem [{\citenamefont {Zhang}\ \emph {et~al.}(2020)\citenamefont {Zhang},
  \citenamefont {He},\ and\ \citenamefont {Ping}}]{Zhang:2020cdi}%
  \BibitemOpen
  \bibfield  {author} {\bibinfo {author} {\bibfnamefont {Q.}~\bibnamefont
  {Zhang}}, \bibinfo {author} {\bibfnamefont {B.-R.}\ \bibnamefont {He}},\ and\
  \bibinfo {author} {\bibfnamefont {J.-L.}\ \bibnamefont {Ping}},\ }\bibfield
  {title} {\bibinfo {title} {{Pentaquarks with the $qqs\bar{Q}Q$ configuration
  in the Chiral Quark Model}},\ }\href@noop {} {\  (\bibinfo {year} {2020})},\
  \Eprint {https://arxiv.org/abs/2006.01042} {arXiv:2006.01042 [hep-ph]}
  \BibitemShut {NoStop}%
\bibitem [{\citenamefont {Paryev}(2024)}]{Paryev:2023uhl}%
  \BibitemOpen
  \bibfield  {author} {\bibinfo {author} {\bibfnamefont {E.~Y.}\ \bibnamefont
  {Paryev}},\ }\bibfield  {title} {\bibinfo {title} {{Probing the hidden-bottom
  pentaquark resonances in photonuclear bottomonium production near threshold:
  Differential observables}},\ }\href
  {https://doi.org/10.1016/j.nuclphysa.2023.122792} {\bibfield  {journal}
  {\bibinfo  {journal} {Nucl. Phys. A}\ }\textbf {\bibinfo {volume} {1042}},\
  \bibinfo {pages} {122792} (\bibinfo {year} {2024})},\ \Eprint
  {https://arxiv.org/abs/2310.04123} {arXiv:2310.04123 [hep-ph]} \BibitemShut
  {NoStop}%
\bibitem [{\citenamefont {Zhang}\ \emph {et~al.}(2023)\citenamefont {Zhang},
  \citenamefont {Liu},\ and\ \citenamefont {Jia}}]{Zhang:2023teh}%
  \BibitemOpen
  \bibfield  {author} {\bibinfo {author} {\bibfnamefont {W.-X.}\ \bibnamefont
  {Zhang}}, \bibinfo {author} {\bibfnamefont {C.-L.}\ \bibnamefont {Liu}},\
  and\ \bibinfo {author} {\bibfnamefont {D.}~\bibnamefont {Jia}},\ }\bibfield
  {title} {\bibinfo {title} {{Study of the hidden-heavy pentaquarks and
  $P_{cs}$ states}},\ }\href@noop {} {\  (\bibinfo {year} {2023})},\ \Eprint
  {https://arxiv.org/abs/2312.12770} {arXiv:2312.12770 [hep-ph]} \BibitemShut
  {NoStop}%
\bibitem [{\citenamefont {Sharma}\ and\ \citenamefont
  {Upadhyay}(2024)}]{Sharma:2024ern}%
  \BibitemOpen
  \bibfield  {author} {\bibinfo {author} {\bibfnamefont {A.}~\bibnamefont
  {Sharma}}\ and\ \bibinfo {author} {\bibfnamefont {A.}~\bibnamefont
  {Upadhyay}},\ }\bibfield  {title} {\bibinfo {title} {{Hidden-Bottom
  Pentaquarks: Mass Spectrum, Magnetic Moments and Partial Widths}},\
  }\href@noop {} {\  (\bibinfo {year} {2024})},\ \Eprint
  {https://arxiv.org/abs/2402.14885} {arXiv:2402.14885 [hep-ph]} \BibitemShut
  {NoStop}%
\bibitem [{\citenamefont {Aaij}\ \emph {et~al.}(2018)\citenamefont {Aaij} \emph
  {et~al.}}]{LHCb:2017wmj}%
  \BibitemOpen
  \bibfield  {author} {\bibinfo {author} {\bibfnamefont {R.}~\bibnamefont
  {Aaij}} \emph {et~al.} (\bibinfo {collaboration} {LHCb}),\ }\bibfield
  {title} {\bibinfo {title} {{Search for weakly decaying $b$-flavored
  pentaquarks}},\ }\href {https://doi.org/10.1103/PhysRevD.97.032010}
  {\bibfield  {journal} {\bibinfo  {journal} {Phys. Rev. D}\ }\textbf {\bibinfo
  {volume} {97}},\ \bibinfo {pages} {032010} (\bibinfo {year} {2018})},\
  \Eprint {https://arxiv.org/abs/1712.08086} {arXiv:1712.08086 [hep-ex]}
  \BibitemShut {NoStop}%
\bibitem [{\citenamefont {\"Ozdem}(2024)}]{Ozdem:2023rkx}%
  \BibitemOpen
  \bibfield  {author} {\bibinfo {author} {\bibfnamefont {U.}~\bibnamefont
  {\"Ozdem}},\ }\bibfield  {title} {\bibinfo {title} {{Analysis of the
  $Z_b(10650)$ state based on electromagnetic properties}},\ }\href
  {https://doi.org/10.1140/epjc/s10052-024-12408-2} {\bibfield  {journal}
  {\bibinfo  {journal} {Eur. Phys. J. C}\ }\textbf {\bibinfo {volume} {84}},\
  \bibinfo {pages} {45} (\bibinfo {year} {2024})},\ \Eprint
  {https://arxiv.org/abs/2311.11327} {arXiv:2311.11327 [hep-ph]} \BibitemShut
  {NoStop}%
\bibitem [{\citenamefont {Lichtenberg}(1977)}]{Lichtenberg:1976fi}%
  \BibitemOpen
  \bibfield  {author} {\bibinfo {author} {\bibfnamefont {D.~B.}\ \bibnamefont
  {Lichtenberg}},\ }\bibfield  {title} {\bibinfo {title} {{Magnetic Moments of
  Charmed Baryons in the Quark Model}},\ }\href
  {https://doi.org/10.1103/PhysRevD.15.345} {\bibfield  {journal} {\bibinfo
  {journal} {Phys. Rev. D}\ }\textbf {\bibinfo {volume} {15}},\ \bibinfo
  {pages} {345} (\bibinfo {year} {1977})}\BibitemShut {NoStop}%
\bibitem [{\citenamefont {Workman}\ \emph {et~al.}(2022)\citenamefont {Workman}
  \emph {et~al.}}]{Workman:2022ynf}%
  \BibitemOpen
  \bibfield  {author} {\bibinfo {author} {\bibfnamefont {R.~L.}\ \bibnamefont
  {Workman}} \emph {et~al.} (\bibinfo {collaboration} {Particle Data Group}),\
  }\bibfield  {title} {\bibinfo {title} {{Review of Particle Physics}},\ }\href
  {https://doi.org/10.1093/ptep/ptac097} {\bibfield  {journal} {\bibinfo
  {journal} {PTEP}\ }\textbf {\bibinfo {volume} {2022}},\ \bibinfo {pages}
  {083C01} (\bibinfo {year} {2022})}\BibitemShut {NoStop}%
\bibitem [{\citenamefont {Ebert}\ \emph {et~al.}(2011)\citenamefont {Ebert},
  \citenamefont {Faustov},\ and\ \citenamefont {Galkin}}]{Ebert:2010af}%
  \BibitemOpen
  \bibfield  {author} {\bibinfo {author} {\bibfnamefont {D.}~\bibnamefont
  {Ebert}}, \bibinfo {author} {\bibfnamefont {R.~N.}\ \bibnamefont {Faustov}},\
  and\ \bibinfo {author} {\bibfnamefont {V.~O.}\ \bibnamefont {Galkin}},\
  }\bibfield  {title} {\bibinfo {title} {{Masses of tetraquarks with open charm
  and bottom}},\ }\href {https://doi.org/10.1016/j.physletb.2010.12.033}
  {\bibfield  {journal} {\bibinfo  {journal} {Phys. Lett. B}\ }\textbf
  {\bibinfo {volume} {696}},\ \bibinfo {pages} {241} (\bibinfo {year}
  {2011})},\ \Eprint {https://arxiv.org/abs/1011.2677} {arXiv:1011.2677
  [hep-ph]} \BibitemShut {NoStop}%
\end{thebibliography}%

\end{document}